\documentclass[journal]{IEEEtran}
\usepackage{graphicx}
\usepackage{amsmath,amssymb,mathtools}
\usepackage{caption}
\usepackage{algorithm}
\usepackage{algpseudocode}
\usepackage{subcaption}
\usepackage{tabularx}
\usepackage{array}
\usepackage{dblfloatfix}
\usepackage{placeins}
\usepackage{orcidlink}
\hypersetup{hidelinks}

\usepackage{cite}
\usepackage{booktabs}
\usepackage{siunitx}
\usepackage{microtype}
\usepackage{float}
\allowdisplaybreaks
\title{Structure-Exploiting ADMM for Low-Complexity
Finite-Blocklength Beamforming in Cellular IoT Networks}
\author{Asim Ihsan \,\orcidlink{0000-0001-7491-7178},
Muhammad Asif \,\orcidlink{0000-0002-9699-1675}, Member, IEEE,
Ali Arshad Nasir\,\orcidlink{0000-0001-5012-1562}, Senior Member, IEEE, Wali Ullah Khan\,\orcidlink{0000-0003-1485-5141}, Galymzhan Nauryzbayev\,\orcidlink{0000-0003-4470-3851}, Senior Member, IEEE, and Khaled M. Rabie\,\orcidlink{0000-0002-9784-3703}, Senior Member, IEEE
\thanks{Corresponding author: Khaled M. Rabie}
\thanks{Asim Ihsan is with the Interdisciplinary Research Center for Communication Systems and Sensing (IRC-CSS); Ali Arshad Nasir is with the Department of Electrical Engineering and the Center for Communication Systems and Sensing; and Khaled M. Rabie is with the Department of Computer Engineering and the Research Center for Communication Systems and Sensing, King Fahd University of Petroleum and Minerals (KFUPM), Dhahran 31261, Saudi Arabia (e-mails: asim.ihsan@kfupm.edu.sa, anasir@kfupm.edu.sa, and k.rabie@kfupm.edu.sa).}
\thanks{Muhammad Asif is with the School of Electrical and Information Engineering, Tongling University, Tongling 244002, China (Email: masif@tlu.edu.cn).}
\thanks{Wali Ullah Khan is with the Interdisciplinary Centre for Security, Reliability, and Trust, University of Luxembourg, 1855 Luxembourg City, Luxembourg (Email: waliullah.khan@uni.lu).}
\thanks{Galymzhan Nauryzbayev is with the Department of Electrical and Computer Engineering, School of Engineering, Nazarbayev University, Kabanbay Batyr 53, Astana 010000, Kazakhstan.}
}
\begin{document}

\bstctlcite{BSTcontrol}

\maketitle

\begin{abstract}
Ultra-reliable and low-latency communication (URLLC)
is essential for mission-critical cellular Internet
of Things (IoT) applications requiring reliable
short-packet transmission and computationally
efficient radio-resource allocation. However,
finite-blocklength (FBL)-aware beamforming becomes
computationally demanding as the numbers of users
and transmit antennas increase. Existing successive
convex approximation (SCA) designs typically rely
on general-purpose convex optimization solvers
without exploiting the algebraic structure of the
resulting convex beamforming problem. This paper develops a structure-exploiting consensus alternating
direction method of multipliers (ADMM) framework
for low-complexity FBL-aware beamforming in downlink
multi-user multiple-input single-output (MU-MISO)
cellular IoT networks. The FBL sum rate is maximized
subject to a total transmit-power constraint and
per-user minimum-rate requirements. Each SCA
iteration yields a convex quadratically constrained
quadratic program (QCQP), reformulated into an
equivalent consensus problem. For an $M$-antenna
base station serving $N$ IoT nodes, the global update
exploits Kronecker structure to replace an
$MN\times MN$ factorization with a shared
$M\times M$ factorization reused across beamformers,
while rank-one channel matrices enable low-complexity
local projections. CVX-SCA serves as a general-purpose
numerical benchmark for solution accuracy and
computational efficiency, whereas standard ADMM-SCA
isolates the gains from structural exploitation.
Numerical results demonstrate close agreement with
CVX-SCA in objective value, feasibility, and
convergence. Under user scaling, the proposed method
achieves more than $4\times$ end-to-end speedup
relative to CVX-SCA and approximately $6\times$
per-iteration speedup relative to standard ADMM-SCA,
highlighting the computational benefits of structural
exploitation for cellular IoT resource allocation.
\end{abstract}
\section{Introduction}
The growing demand for mission-critical Internet of Things (IoT) applications, including industrial automation, intelligent transportation, remote healthcare, and real-time control, has intensified the need for reliable and low-latency wireless connectivity. Ultra-reliable and low-latency communication (URLLC), a fundamental service category of fifth-generation (5G) and beyond
cellular networks, provides an enabling framework for
supporting such latency-critical IoT applications~\cite{9726784,10918329}.
Unlike enhanced mobile broadband (eMBB), URLLC
requires extremely stringent quality-of-service (QoS)
guarantees in terms of both transmission latency and
communication reliability, where end-to-end latencies
on the order of milliseconds and packet error
probabilities as low as $10^{-5}$ or below are
typically required~\cite{9246274,8472907}.
Such stringent performance requirements fundamentally
change the physical-layer transmission design and
resource allocation strategies adopted in conventional wireless communication systems.

To satisfy these stringent latency requirements, URLLC
systems rely on short-packet transmission, where only a
finite number of channel uses are available for reliable data delivery. Consequently, the classical Shannon-capacity
formula, which assumes infinitely long codewords and
vanishing decoding error probability, is no longer applicable. Instead, finite blocklength (FBL) information theory explicitly characterizes the fundamental tradeoff among transmission rate, decoding reliability, and blocklength, providing a more accurate performance metric for practical URLLC systems~\cite{5452208,7529226}. Compared
with conventional Shannon-capacity formulations, the FBL achievable-rate expression introduces an additional channel-dispersion penalty that depends nonlinearly on the instantaneous signal-to-interference-plus-noise ratio (SINR). This dependence makes FBL-aware resource allocation
and beamforming substantially more challenging in
multi-user cellular IoT networks, where interference
coupling and per-user QoS requirements must be
jointly addressed.

Multi-user multiple-input single-output (MU-MISO) beamforming is an effective physical-layer technique for improving spectral efficiency and communication reliability through spatial multiplexing and interference management~\cite{9930879,10430085}. Consequently, FBL-aware active beamforming has received increasing research attention. For conventional MU-MISO transmission, max--min FBL-rate optimization was investigated in~\cite{9246274}, whereas weighted FBL sum-rate, energy-efficiency, and user-fairness designs were considered in~\cite{9467811}. Robust FBL beamforming under imperfect channel state information was subsequently studied in~\cite{9893403}. FBL transmission has also been investigated for more advanced system architectures and multiple-access strategies. Multi-RIS-assisted MU-MISO beamforming and resource allocation were addressed in~\cite{9939154}, while cell-free massive-MIMO designs were considered in~\cite{9399808}. For rate-splitting multiple access (RSMA), FBL sum-rate maximization through joint common and private precoder design was investigated in~\cite{9831048}, whereas an FBL max--min rate design was developed in~\cite{9991090}. More recent studies have examined FBL multicast beamforming for massive-MIMO systems~\cite{10599218}, hybrid beamforming for millimeter-wave massive-MIMO systems~\cite{10839437}, and beamforming designs that directly minimize the FBL decoding error probability~\cite{10769474,11173679}. Joint active beamforming and UAV trajectory optimization has also been considered under an aggregate FBL-rate objective with mobility and propulsion penalties~\cite{ihsan2026online}.

Despite the substantial progress in FBL-aware beamforming, the
computational burden of iterative optimization remains an important
challenge. Existing designs commonly address the underlying
nonconvexity through path-following, successive convex approximation
(SCA), semidefinite relaxation, and related techniques
~\cite{9246274,9467811,9893403,9399808}, resulting in a sequence of tractable convex approximations.
Each approximation must, however, be solved numerically at the corresponding outer iteration, typically using general-purpose convex optimization methods. For many convex and conic formulations, the underlying numerical solution relies on interior-point methods~\cite{vandenberghe2004convex}, whose computational cost can
grow rapidly with the problem dimension. Moreover, general-purpose
numerical methods are designed for broad classes of optimization
problems and do not necessarily exploit the recurring algebraic
structure of the convexified FBL beamforming subproblem. The repeated inner optimization can therefore become a significant computational bottleneck for latency-critical cellular IoT networks, particularly as the numbers of simultaneously served users and transmit antennas increase.

The above limitation motivates a problem-specific treatment of the
inner optimization rather than another approximation of the FBL
rate. In this work, the full active-beamforming formulation and the
outer SCA framework are retained, while the resulting convex beamforming problem
is solved using a dedicated first-order method.
Among such methods, the alternating direction method of multipliers
(ADMM) is particularly relevant because its variable-splitting
mechanism enables coupled optimization problems to be decomposed
into simpler subproblems with alternating updates~\cite{6892987,7889039}. Consensus ADMM has been developed for general quadratically
constrained quadratic programs (QCQPs)~\cite{7517329}, and ADMM-based decomposition has been
applied to robust multicast, RIS-assisted distributed, and coordinated multi-cell hybrid
beamforming~\cite{9736691,10210090,9921303}. These methods, however, are designed for non-FBL objectives and
different system architectures, and their decompositions do not carry
over to the FBL sum-rate-maximizing transmit-beamforming problem. In particular, the channel-dispersion term modifies both the SCA objective and the per-user rate constraints, while the
adopted approximation introduces additional trust-region constraints over the jointly coupled transmit beamformers. More importantly, they do
not exploit the particular global and local matrix structures that arise in the resulting convex beamforming problem. This leaves open the design of an ADMM-based inner solver whose updates are customized specifically to
the structure of FBL MU-MISO beamforming. Although ADMM has appeared in the FBL literature
~\cite{song2023delay}, its variable-splitting mechanism is employed
for RIS phase-shift optimization rather than active transmit
beamforming. The receive beamformer is obtained analytically using
an MMSE receiver, while the RIS phase-shift QCQP is handled by a
general-purpose convex solver. Hence, a dedicated
structure-exploiting ADMM solver for full FBL transmit-beamforming
optimization remains unaddressed.

A direct application of consensus ADMM does not, however,
automatically provide the desired computational reduction. For a
transmitter with $M$ antennas serving $N$ users, stacking the $N$
beamforming vectors yields an $MN$-dimensional optimization
variable, such that a standard implementation retains
full-dimensional operations in both the global update and the local
QCQP projections. The global step may require the factorization of
an $MN\times MN$ linear system, while generic local projections do
not exploit the rank-one structure induced by the user channels.
Hence, the computational advantage depends not only on adopting a
first-order decomposition, but also on exploiting the algebraic
structure of the resulting ADMM updates. The proposed framework addresses both computational bottlenecks by
exploiting the structure of the convex beamforming problem in the
global and local ADMM updates. In the global update, the quadratic
coefficient matrix exhibits a repeated Kronecker structure across
the $N$ beamforming blocks. Rather than factorizing the resulting
$MN\times MN$ system, this structure reduces the update to a single
$M\times M$ factorization that is reused for all users. In the local
updates, the user-dependent quadratic constraints inherit rank-one
matrices from the channel outer products. This property enables the
corresponding inverse actions to be evaluated directly through the
Sherman--Morrison formula, avoiding the explicit formation or
inversion of high-dimensional matrices. Hence, the proposed method
retains the full active-beamforming formulation and the underlying
consensus problem while reducing the computational cost of both the
global and local ADMM operations. To the best of the authors'
knowledge, such a structure-exploiting consensus-ADMM inner solver
has not been developed for FBL sum-rate-maximizing MU-MISO
beamforming.

\begin{table*}[!t]
\centering
\caption{Comparison With Representative Rate-Oriented FBL Active-Beamforming Frameworks}
\label{tab}
\footnotesize
\setlength{\tabcolsep}{1.5pt}
\renewcommand{\arraystretch}{1.18}

\begin{tabularx}{\textwidth}{
@{}
>{\hsize=.90\hsize\linewidth=\hsize\raggedright\arraybackslash}X
>{\hsize=1.05\hsize\linewidth=\hsize\raggedright\arraybackslash}X
>{\hsize=1.05\hsize\linewidth=\hsize\raggedright\arraybackslash}X
>{\centering\arraybackslash}p{2.00cm}
>{\centering\arraybackslash}p{1.90cm}
>{\centering\arraybackslash}p{2.20cm}
>{\centering\arraybackslash}p{2.65cm}
@{}
}
\toprule

\textbf{Work}
&
\textbf{System/Access}
&
\textbf{FBL Rate Objective}
&
\shortstack{\textbf{Active BF}\\\textbf{Optimization}}
&
\shortstack{\textbf{SCA-Based}\\\textbf{Design}}
&
\shortstack{\textbf{First-Order}\\\textbf{Inner Solver}}
&
\shortstack{\textbf{Structure-}\\\textbf{Exploiting Solver}}
\\

\midrule

Nasir \emph{et al.}~\cite{9246274}
&
MU-MISO/SDMA
&
Max--min FBL rate
&
$\checkmark$
&
$\checkmark$
&
$\times$
&
$\times$
\\

He \emph{et al.}~\cite{9467811}
&
MU-MISO/SDMA
&
Weighted FBL sum rate
&
$\checkmark$
&
$\checkmark$
&
$\times$
&
$\times$
\\

Xu \emph{et al.}~\cite{9831048}
&
MU-MISO/RSMA
&
FBL sum rate
&
$\checkmark$
&
$\checkmark$
&
$\times$
&
$\times$
\\

Xu \emph{et al.}~\cite{9991090}
&
MU-MISO/RSMA
&
Max--min FBL rate
&
$\checkmark$
&
$\checkmark$
&
$\times$
&
$\times$
\\

Ihsan \emph{et al.}~\cite{ihsan2026online}
&
UAV MU-MISO/SDMA
&
FBL sum rate 
&
$\checkmark$
&
$\checkmark$
&
$\times$
&
$\times$
\\

\midrule

\textbf{This work}
&
\textbf{MU-MISO/SDMA}
&
\textbf{FBL sum rate}
&
$\boldsymbol{\checkmark}$
&
$\boldsymbol{\checkmark}$
&
$\boldsymbol{\checkmark}$
&
$\boldsymbol{\checkmark}$
\\

\bottomrule
\end{tabularx}
\end{table*}

Table~\ref{tab} summarizes the computational positioning of the
proposed framework relative to representative rate-oriented FBL
active-beamforming methods. In particular, the comparison
distinguishes the established use of active beamforming and
successive convexification from the first-order and
structure-exploiting treatment of the resulting inner optimization
problem. Motivated by the above gap, this paper develops a structure-aware
consensus ADMM-SCA framework for FBL sum-rate maximization in
downlink MU-MISO URLLC systems subject to a total transmit-power
constraint and per-user minimum FBL-rate requirements. The outer SCA framework is retained, while each convex
full-beamforming QCQP is cast into an equivalent consensus form and
solved using the proposed structure-aware first-order implementation.

The main contributions of this paper are summarized as follows.

\begin{itemize}

\item We develop a structure-exploiting consensus ADMM-SCA framework for FBL sum-rate-maximizing active
beamforming in downlink MU-MISO cellular IoT networks
supporting URLLC, subject to a total transmit-power
constraint and per-user minimum-rate requirements.
 At each outer SCA iteration, the resulting convex beamforming QCQP is cast into an equivalent consensus form and solved using a dedicated first-order inner algorithm. The consensus decomposition
separates the coupled beamforming design into a global update and independent constraint-wise local projections that can be executed in parallel.

\item We reveal and exploit the hidden Kronecker structure of the aggregate quadratic objective. The proposed implementation reduces the global ADMM update from factoring an $MN\times MN$ matrix to factoring a shared $M\times M$ matrix and reusing the resulting factorization across all $N$ beamforming blocks. This reduction avoids matrix-factorization operations on the full beamforming dimension.

\item We derive a unified low-complexity projection procedure for the local QoS and trust-region constraints. By expressing these constraints in a common QCQP form and exploiting
the rank-one structure induced by the user channel matrices, the
Sherman--Morrison inverse action is evaluated directly through inner
products and scaled vector operations. The different local
constraints can therefore be handled by the same projection routine
without explicitly forming or inverting high-dimensional matrices.

\item We characterize the computational complexity and scalability of the proposed implementation relative to both standard consensus ADMM and CVX/MOSEK. The analysis establishes that the computational advantage arises from exploiting the Kronecker and rank-one structures of the FBL beamforming problem rather than from the use of ADMM alone. The measured per-iteration speedup over standard ADMM reaches
approximately $6\times$ for $N=12$ and $1.9\times$ for $M=24$,
demonstrating that the benefit of structural exploitation increases
relative to standard ADMM as either the number of users or transmit
antennas increases.

\item We provide extensive numerical validation of solution accuracy, convergence, and end-to-end computational efficiency. The proposed structured ADMM-SCA framework closely reproduces the objective values, constraint satisfaction, and outer-SCA convergence behavior of the CVX/MOSEK-based implementation while requiring substantially less execution time. Agreement across independent channel realizations and the observed primal--dual residual convergence further validate the reliability of the proposed inner solver. The numerical results also characterize the effects of blocklength, decoding reliability, antenna resources, and QoS requirements on the achievable FBL sum rate.

\end{itemize}

The remainder of this paper is organized as follows. Section~II presents the downlink MU-MISO cellular IoT
system model. Section~III formulates the FBL sum-rate maximization problem and develops its SCA-based convex reformulation. Section~IV presents the stacked matrix and consensus representations
of the convex beamforming QCQP, derives the proposed
structure-exploiting ADMM solver, and analyzes its computational
complexity. Section~V presents the numerical results, and Section~VI concludes the paper.

\section{System Model}
\begin{figure}[!t]
    \centering
    \includegraphics[width=\linewidth]
    {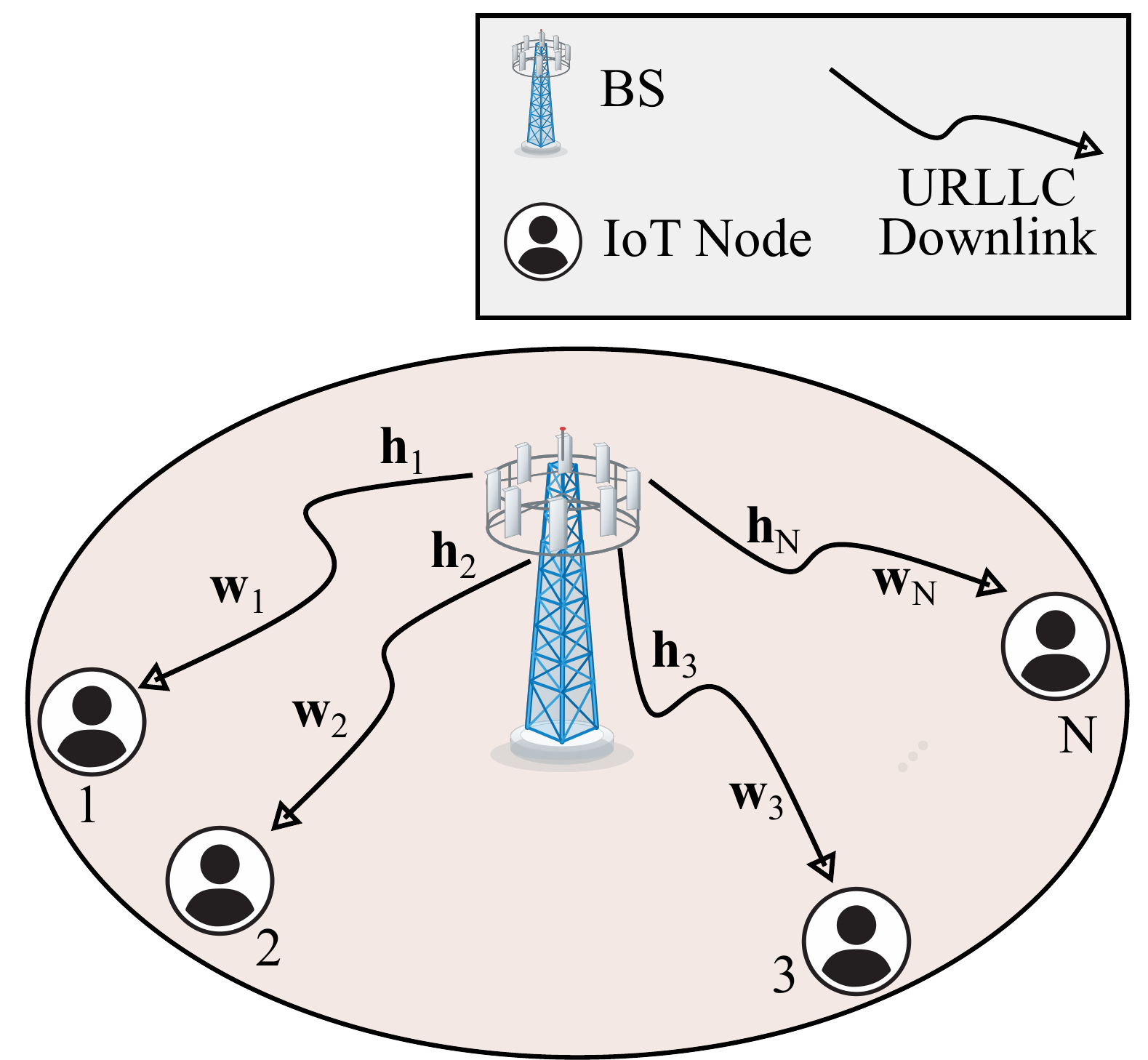}
    \caption{Downlink cellular IoT network supporting
URLLC through MU-MISO beamforming.}
    \label{fig:system_model}
\end{figure}

Consider the downlink cellular IoT network supporting
URLLC services illustrated in Fig.~\ref{fig:system_model}, where a base station (BS) equipped with $M$ transmit antennas simultaneously serves $N$ single-antenna
latency-critical IoT nodes over a common time-frequency resource block. These nodes may represent industrial sensing and control devices requiring reliable short-packet communication. Let
$\mathcal{N}=\{1,\ldots,N\}$ denote the user set.

The transmitted signal is given by
\begin{equation}
\mathbf{u}
=
\sum_{n\in\mathcal N}
\mathbf{w}_{n}s_{n},
\label{eq:tx_signal}
\end{equation}
where $\mathbf{w}_{n}\in\mathbb{C}^{M}$ denotes the beamforming
vector intended for user $n$, and $s_n$ is the corresponding
information symbol. The information symbols are assumed to be
mutually independent and normalized such that
$\mathbb{E}\{s_ns_k^{*}\}=\delta_{nk}$, where $\delta_{nk}$ denotes
the Kronecker delta.

Following \cite{9246274}, a quasi-static flat-fading channel is
considered, where the channel remains constant over each transmission
block. Let $\mathbf{h}_n\in\mathbb{C}^{M}$ denote the channel vector
from the BS to user $n$. The BS is assumed to have perfect
instantaneous channel state information (CSI) of all users. The signal
received by user $n$ is
\begin{equation}
y_n
=
\mathbf{h}_n^H\mathbf{w}_ns_n
+
\sum_{\substack{k\in\mathcal N\\k\neq n}}
\mathbf{h}_n^H\mathbf{w}_ks_k
+
\nu_n,
\label{eq:received_signal}
\end{equation}
where $\nu_n\sim\mathcal{CN}(0,\sigma^2)$ denotes additive white
Gaussian noise.

By treating multiuser interference as noise, the received
signal-to-interference-plus-noise ratio (SINR) of user $n$ is
\begin{equation}
\gamma_n
=
\frac{S_n}{I_n},
\label{eq:sinr}
\end{equation}
where
\begin{align}
S_n
&=
\left|
\mathbf{h}_n^H\mathbf{w}_n
\right|^2,
\label{eq:signal_power}
\\
I_n
&=
\sum_{\substack{k\in\mathcal N\\k\neq n}}
\left|
\mathbf{h}_n^H\mathbf{w}_k
\right|^2
+
\sigma^2.
\label{eq:interference_power}
\end{align}

Under finite-blocklength (FBL) transmission, the achievable rate
of user $n$ is approximated using the normal approximation
\cite{5452208} as
\begin{equation}
R_n
\approx
\log(1+\gamma_n)
-
\frac{Q^{-1}(\varepsilon)}{\sqrt{L}}
\sqrt{V_n},
\label{eq:fbl_rate}
\end{equation}
where $L$ denotes the channel blocklength, $\varepsilon$ is the
target packet decoding error probability, and $Q^{-1}(\cdot)$ is
the inverse Gaussian $Q$-function. Moreover,
\begin{equation}
V_n
=
1-
\frac{1}{(1+\gamma_n)^2}
\label{eq:dispersion}
\end{equation}
denotes the channel dispersion. Throughout this paper,
$\log(\cdot)$ denotes the natural logarithm; hence, $R_n$ is
measured in nats/s/Hz.

For the subsequent optimization development, we adopt the normal
approximation in \eqref{eq:fbl_rate} as the FBL rate model and write
\begin{equation}
R_n
=
C_n-D_n,
\label{eq:rate_decomposition}
\end{equation}
where
\begin{align}
C_n
&=
\log(1+\gamma_n),
\label{eq:capacity}
\\
D_n
&=
c\sqrt{V_n},
\label{eq:dispersion_penalty}
\end{align}
with $c=Q^{-1}(\epsilon)/\sqrt{L}$.

\section{Convex Reformulation of the FBL Beamforming Problem}

Based on the system model, the FBL beamforming design is
formulated as
\begin{subequations}
\label{P1}
\begin{align}
(\mathcal{P}_1):\quad
\underset{\{\mathbf{w}_n\}_{n\in\mathcal N}}{\mathrm{maximize}}
\quad &
\sum_{n\in\mathcal N} R_n
\label{P1_obj}
\\
\mathrm{s.t.}\quad
&
\frac{1}{\eta}
\sum_{n\in\mathcal N}
\|\mathbf{w}_n\|_2^2
\le
P_{\max},
\label{P1_power}
\\
&
R_n
\ge
R_n^{\min},
\quad
\forall n\in\mathcal N.
\label{P1_rate}
\end{align}
\end{subequations}
where $P_{\max}$ denotes the maximum DC power budget and
$\eta\in(0,1]$ denotes the power-amplifier efficiency. Problem~\eqref{P1} is nonconvex because the beamforming
vectors are coupled through the SINR expressions appearing in
both the Shannon-rate and FBL dispersion terms. SCA provides a general framework for handling nonconvex optimization
problems through a sequence of tractable convex approximations
~\cite{7776948}. Using the SCA reformulation in~\cite{ihsan2026online},
problem~\eqref{P1} is successively approximated by a
sequence of convex beamforming subproblems. The initialization
and surrogate construction are presented next.
\subsection{Initialization and Successive Convex Approximation}

Before starting the SCA iterations, a power-feasible
beamforming warm start is generated using the max--min
Shannon-rate initialization of~\cite{ihsan2026online}.
Specifically, the initial beamforming vectors are obtained by
approximately solving
\begin{equation*}
\begin{aligned}
(\mathrm{P}_{\mathrm{Init}}):\quad
\underset{\{\mathbf{w}_n\}_{n\in\mathcal N}}
{\operatorname{maximize}}
\quad & \min_{n\in\mathcal N}\log(1+\gamma_n)\\
\operatorname{subject~to}\quad
& \frac{1}{\eta}
\sum_{n\in\mathcal N}\|\mathbf{w}_n\|_2^2
\leq P_{\max}.
\end{aligned}
\end{equation*}
The resulting beamforming vectors are used as the initial SCA
expansion point
$\{\mathbf{w}_n^{(0)}\}_{n\in\mathcal N}$. Let $\{\mathbf{w}_n^{(i)}\}_{n\in\mathcal N}$ denote the
beamforming vectors available at SCA iteration $i$. Define
\begin{align}
\chi_n^{(i)}
&=
\mathbf h_n^H\mathbf w_n^{(i)},
\label{eq:chi_desired}
\\
\chi_{k,n}^{(i)}
&=
\mathbf h_n^H\mathbf w_k^{(i)},
\quad k\neq n,
\label{eq:chi_interference}
\\
S_n^{(i)}
&=
\left|\chi_n^{(i)}\right|^2,
\label{eq:Siter}
\\
I_n^{(i)}
&=
\sum_{\substack{k\in\mathcal N\\k\neq n}}
\left|\chi_{k,n}^{(i)}\right|^2
+
\sigma^2,
\label{eq:Iiter}
\\
\gamma_n^{(i)}
&=
\frac{S_n^{(i)}}{I_n^{(i)}},
\label{eq:gamma_iter}
\\
V_n^{(i)}
&=
1-
\frac{1}
{\left(1+\gamma_n^{(i)}\right)^2}.
\label{eq:Viter}
\end{align}

At iteration $i$, the Shannon-rate and FBL dispersion terms
are replaced by a locally tight concave lower surrogate and a
locally tight convex upper surrogate, respectively. The
concave surrogate of the Shannon-rate term is
\begin{equation}
\bar C_n^{(i)}
=
C_n^{(i)}
+
L_n^{(i)},
\label{eq:Cbar}
\end{equation}
where
\begin{align}
C_n^{(i)}
&=
\log\left(1+\gamma_n^{(i)}\right),
\label{eq:Citer}
\\
L_n^{(i)}
&=
\frac{
2\Re\left\{
(\chi_n^{(i)})^{*}
\mathbf h_n^H\mathbf w_n
\right\}
}{
I_n^{(i)}
}
-
\gamma_n^{(i)}
-
\eta_n^{(i)}
\left(S_n+I_n\right),
\label{eq:Liter}
\\
\eta_n^{(i)}
&=
\frac{
S_n^{(i)}
}{
I_n^{(i)}
\left(S_n^{(i)}+I_n^{(i)}\right)
}.
\label{eq:eta_iter}
\end{align}

The convex upper surrogate of the FBL dispersion penalty is
\begin{equation}
\bar D_n^{(i)}
=
c\left(
B_n^{(i)}
-
A_n^{(i)}
\Xi_n^{(i)}
\right),
\label{eq:Dbar}
\end{equation}
where
\begin{align}
A_n^{(i)}
&=
\frac{1}{2\sqrt{V_n^{(i)}}},
\label{eq:Aiter}
\\
B_n^{(i)}
&=
\frac{\sqrt{V_n^{(i)}}}{2}
+
A_n^{(i)},
\label{eq:Biter}
\\
\Xi_n^{(i)}
&=
\alpha_n^{(i)}\bar I_n^{(i)}
-
\beta_n^{(i)}
\left(S_n+I_n\right)
-
\psi_n^{(i)},
\label{eq:Xiiter}
\end{align}
with
\begin{align}
\alpha_n^{(i)}
&=
\frac{
4I_n^{(i)}
}{
\left(S_n^{(i)}+I_n^{(i)}\right)^2
},
\label{eq:alpha_iter}
\\
\beta_n^{(i)}
&=
\frac{
2\left(I_n^{(i)}\right)^2
}{
\left(S_n^{(i)}+I_n^{(i)}\right)^3
},
\label{eq:beta_iter}
\\
\psi_n^{(i)}
&=
\frac{
\left(I_n^{(i)}\right)^2
}{
\left(S_n^{(i)}+I_n^{(i)}\right)^2
}.
\label{eq:psi_iter}
\end{align}

The affine lower approximation of the interference-plus-noise
power is given by
\begin{equation}
\bar{I}_n^{(i)}
=
\sum_{\substack{k\in\mathcal N\\k\neq n}}
\left[
2\Re
\left\{
(\chi_{k,n}^{(i)})^{*}
\mathbf{h}_n^H\mathbf{w}_k
\right\}
-
|\chi_{k,n}^{(i)}|^2
\right]
+
\sigma^2.
\label{eq:Ibar}
\end{equation}

Combining \eqref{eq:Cbar} and \eqref{eq:Dbar}, the concave
surrogate of the FBL achievable rate is
\begin{equation}
\bar R_n^{(i)}
=
\bar C_n^{(i)}
-
\bar D_n^{(i)}.
\label{eq:Rbar}
\end{equation}
The validity of the dispersion surrogate in \eqref{eq:Dbar}
requires restricting the optimization variables to the SCA
trust region adopted in \cite{ihsan2026online}. The corresponding
trust-region constraints are included explicitly in the convex
subproblem below.

\subsection{Convex Beamforming Problem}

Substituting the surrogate functions \eqref{eq:Cbar} and
\eqref{eq:Dbar} into Problem~\eqref{P1} yields the following
convex approximation at the $i$-th SCA iteration:

\begin{subequations}
\label{P2}
\begin{align}
(\mathcal{P}_2^{(i)}):\quad
&
\underset{\{\mathbf{w}_n\}_{n\in\mathcal N}}{\mathrm{maximize}}
\quad
\sum_{n\in\mathcal N}\bar{R}_n^{(i)}
\label{P2_obj}
\\
\mathrm{s.t.}\quad
&
\sum_{n\in\mathcal N}
\|\mathbf{w}_n\|_2^2
\le
\eta P_{\max},
\label{P2_power}
\\
&
\bar{R}_n^{(i)}
\ge
R_n^{\min},
\quad
\forall n\in\mathcal N,
\label{P2_rate}
\\
&
S_n+I_n
\le
2\left(
S_n^{(i)}+I_n^{(i)}
\right),
\quad
\forall n\in\mathcal N,
\label{P2_tr1}
\\
&
\frac{S_n+I_n}
{S_n^{(i)}+I_n^{(i)}}
\le
\frac{2\bar I_n^{(i)}}
{I_n^{(i)}},
\quad
\forall n\in\mathcal N.
\label{P2_tr2}
\end{align}
\end{subequations}

Constraints~\eqref{P2_tr1} and \eqref{P2_tr2} enforce the SCA
trust region described above. Problem~\eqref{P2} is convex
because $\bar R_n$ is concave in the beamforming vectors and
all the constraints define convex feasible sets. Although
Problem~\eqref{P2} can be solved using generic interior-point
solvers, repeatedly solving it at every SCA iteration may incur
high computational complexity, particularly in large-scale
MU-MISO systems. Therefore, the following section develops a
structure-exploiting consensus ADMM algorithm for efficiently
solving Problem~\eqref{P2}.

\section{Proposed Consensus ADMM Beamforming Algorithm}
This section develops a consensus alternating direction method of
multipliers (ADMM) algorithm for solving the convex QCQP
in~\eqref{P2}. To enable consensus decomposition, the individual
beamforming vectors are collected into an aggregate variable, and
the quadratic objective and constraints are written in
stacked matrix form. This representation makes their coefficient
matrices explicit and reveals the algebraic structures exploited in
the global and local ADMM updates.

\subsection{Stacked Matrix Representation of the QCQP}

Consider the stacked beamforming vector
\begin{equation}
\mathbf{x}
=
\operatorname{vec}
\left(
\left[
\mathbf{w}_1,
\mathbf{w}_2,
\ldots,
\mathbf{w}_N
\right]
\right)
\in
\mathbb{C}^{MN},
\label{eq:x}
\end{equation}
where $\operatorname{vec}(\cdot)$ denotes the column-wise
vectorization operator.

For user $n$, define
$\mathbf H_n=\mathbf h_n\mathbf h_n^H$, together with
\begin{align}
\widehat{\mathbf H}_n
&=
\mathbf I_N
\otimes
\mathbf H_n,
\label{eq:Hhat}
\\
\widetilde{\mathbf H}_n
&=
\mathbf E_n
\otimes
\mathbf H_n,
\label{eq:Htilde}
\end{align}
where $\mathbf E_n=\mathbf e_n\mathbf e_n^T$ and
$\mathbf e_n$ denotes the $n$th canonical basis vector of
$\mathbb R^N$. The matrix $\widehat{\mathbf H}_n$ captures the
contributions of all transmitted beams to user $n$, whereas
$\widetilde{\mathbf H}_n$ isolates the contribution of its intended
beam. The aggregate interference matrix is then defined as
\begin{equation}
\mathbf H_{I,n}
=
\widehat{\mathbf H}_n
-
\widetilde{\mathbf H}_n.
\label{eq:HI}
\end{equation}

Using \eqref{eq:Hhat}--\eqref{eq:HI}, the desired signal and the
interference-plus-noise powers are equivalently written as
\begin{align}
S_n
&=
\mathbf x^H
\widetilde{\mathbf H}_n
\mathbf x,
\label{eq:Squad}
\\
S_n+I_n
&=
\mathbf x^H
\widehat{\mathbf H}_n
\mathbf x
+
\sigma^2,
\label{eq:SIquad}
\\
I_n
&=
\mathbf x^H
\mathbf H_{I,n}
\mathbf x
+
\sigma^2.
\label{eq:Iquad}
\end{align}

Furthermore, the affine approximation of the interference term
introduced in Section~III becomes
\begin{equation}
\bar I_n^{(i)}
=
2\Re
\left\{
(\mathbf g_n^{(i)})^H\mathbf x
\right\}
+
\delta_n^{(i)},
\label{eq:Ilinear}
\end{equation}
where
\begin{align}
\mathbf g_n^{(i)}
&=
\mathbf H_{I,n}\mathbf x^{(i)},
\label{eq:g}
\\
\delta_n^{(i)}
&=
\sigma^2
-
(\mathbf x^{(i)})^H
\mathbf H_{I,n}
\mathbf x^{(i)}.
\label{eq:delta}
\end{align}

Using the matrix representations in
\eqref{eq:Squad}--\eqref{eq:Ilinear}, the surrogate functions
derived in Section~III admit the following canonical matrix
representations with respect to the stacked beamforming vector
$\mathbf{x}$. Specifically, the Shannon-rate surrogate in \eqref{eq:Cbar} is
equivalently written as
\begin{equation}
\bar C_n^{(i)}
=
\kappa_n^{\mathrm C,(i)}
+
2\Re
\left\{
(\mathbf f_n^{\mathrm C,(i)})^H\mathbf x
\right\}
-
\mathbf x^H
\mathbf Q_n^{\mathrm C,(i)}
\mathbf x,
\label{eq:CQCQP}
\end{equation}
where
\begin{align}
\mathbf Q_n^{\mathrm C,(i)}
&=
\eta_n^{(i)}
\widehat{\mathbf H}_n,
\label{eq:QC}
\\
\mathbf f_n^{\mathrm C,(i)}
&=
\frac{
\widetilde{\mathbf H}_n\mathbf x^{(i)}
}{
I_n^{(i)}
},
\label{eq:fC}
\\
\kappa_n^{\mathrm C,(i)}
&=
C_n^{(i)}
-
\gamma_n^{(i)}
-
\eta_n^{(i)}\sigma^2.
\label{eq:kappaC}
\end{align}
with
\begin{equation}
\eta_n^{(i)}
=
\frac{
S_n^{(i)}
}
{
I_n^{(i)}
\left(
S_n^{(i)}
+
I_n^{(i)}
\right)
}.
\label{eq:etaQC}
\end{equation}

Similarly, the FBL dispersion surrogate in \eqref{eq:Dbar}
admits the quadratic representation
\begin{equation}
\bar D_n^{(i)}
=
\kappa_n^{\mathrm D,(i)}
+
2\Re
\left\{
(\mathbf f_n^{\mathrm D,(i)})^H\mathbf x
\right\}
+
\mathbf x^H
\mathbf Q_n^{\mathrm D,(i)}
\mathbf x,
\label{eq:DQCQP}
\end{equation}
where
\begin{align}
&\mathbf Q_n^{\mathrm D,(i)}
=
cA_n^{(i)}
\beta_n^{(i)}
\widehat{\mathbf H}_n,
\label{eq:QD}
\\
&\mathbf f_n^{\mathrm D,(i)}
=
-
cA_n^{(i)}
\alpha_n^{(i)}
\mathbf g_n^{(i)},
\label{eq:fD}
\\
&\kappa_n^{\mathrm D,(i)}
=
cB_n^{(i)}
-
cA_n^{(i)}
\alpha_n^{(i)}
\delta_n^{(i)}
+
cA_n^{(i)}
\beta_n^{(i)}
\sigma^2
+
cA_n^{(i)}
\psi_n^{(i)}.
\label{eq:kappaD}
\end{align}

Substituting
\eqref{eq:CQCQP}
and
\eqref{eq:DQCQP}
into
\eqref{eq:Rbar}
gives
\begin{equation}
\bar R_n^{(i)}
=
\kappa_n^{(i)}
+
2\Re
\left\{
(\mathbf f_n^{(i)})^H\mathbf x
\right\}
-
\mathbf x^H
\mathbf Q_n^{(i)}
\mathbf x,
\label{eq:RQCQP}
\end{equation}
where
\begin{align}
\mathbf Q_n^{(i)}
&=
\mathbf Q_n^{\mathrm C,(i)}
+
\mathbf Q_n^{\mathrm D,(i)},
\label{eq:QFinal}
\\
\mathbf f_n^{(i)}
&=
\mathbf f_n^{\mathrm C,(i)}
-
\mathbf f_n^{\mathrm D,(i)},
\label{eq:fFinal}
\\
\kappa_n^{(i)}
&=
\kappa_n^{\mathrm C,(i)}
-
\kappa_n^{\mathrm D,(i)}.
\label{eq:kappaFinal}
\end{align}

Summing \eqref{eq:RQCQP} over all users yields the aggregate
surrogate objective
\begin{equation}
\sum_{n\in\mathcal N}\bar R_n^{(i)}
=
\kappa^{(i)}
+
2\Re
\left\{
(\mathbf f^{(i)})^H\mathbf x
\right\}
-
\mathbf x^H
\mathbf Q^{(i)}
\mathbf x,
\label{eq:GlobalObjective}
\end{equation}
where
\begin{align}
\mathbf Q^{(i)}
&=
\sum_{n\in\mathcal N}\mathbf Q_n^{(i)},
\label{eq:GlobalQ}
\\
\mathbf f^{(i)}
&=
\sum_{n\in\mathcal N}\mathbf f_n^{(i)},
\label{eq:Globalf}
\\
\kappa^{(i)}
&=
\sum_{n\in\mathcal N}\kappa_n^{(i)}.
\label{eq:Globalkappa}
\end{align}

The stacked matrix representation of the objective extends directly
to the constraints in Problem~\eqref{P2}. In particular, the QoS,
transmit-power, and trust-region constraints can be written in
canonical matrix form with respect to the stacked beamforming vector
$\mathbf{x}$.

The QoS constraint in \eqref{P2_rate} is equivalently written as
\begin{equation}
\mathbf x^H
\mathbf Q_n^{(i)}
\mathbf x
-
2\Re
\left\{
(\mathbf f_n^{(i)})^H\mathbf x
\right\}
\le
\kappa_n^{(i)}
-
R_n^{\min},
\quad
\forall n\in\mathcal N.
\label{eq:QoSQCQP}
\end{equation}
where $\mathbf Q_n^{(i)}$, $\mathbf f_n^{(i)}$, and
$\kappa_n^{(i)}$ are defined in
\eqref{eq:QFinal}--\eqref{eq:kappaFinal}.

Similarly, the transmit-power constraint is represented as
\begin{equation}
\mathbf{x}^{H}\mathbf{x}
\le
\eta P_{\max},
\label{eq:PowerQCQP}
\end{equation}
which is a convex quadratic constraint.

Furthermore, the first trust-region constraint becomes
\begin{equation}
\mathbf{x}^{H}
\widehat{\mathbf H}_{n}
\mathbf{x}
\le
2
\left(
S_{n}^{(i)}
+
I_{n}^{(i)}
\right)
-
\sigma^{2},
\quad
\forall n\in\mathcal N,
\label{eq:TR1QCQP}
\end{equation}
whereas the second trust-region constraint is expressed as
\begin{equation}
\mathbf x^H
\mathbf Q_n^{\mathrm{TR2},(i)}
\mathbf x
-
2\Re
\left\{
(\mathbf f_n^{\mathrm{TR2},(i)})^H
\mathbf x
\right\}
\le
c_n^{\mathrm{TR2},(i)},
\quad
\forall n\in\mathcal N.
\label{eq:TR2QCQP}
\end{equation}
where
\begin{align}
\mathbf Q_n^{\mathrm{TR2},(i)}
&=
\frac{
\widehat{\mathbf H}_n
}{
S_n^{(i)}+I_n^{(i)}
},
\\
\mathbf f_n^{\mathrm{TR2},(i)}
&=
\frac{
2\mathbf g_n^{(i)}
}{
I_n^{(i)}
},
\\
c_n^{\mathrm{TR2},(i)}
&=
\frac{
2\delta_n^{(i)}
}{
I_n^{(i)}
}
-
\frac{
\sigma^2
}{
S_n^{(i)}+I_n^{(i)}
}.
\end{align}

Accordingly, the convex beamforming problem in
\eqref{P2}
is equivalently reformulated as
\begin{subequations}
\label{P3}
\begin{align}
&(\mathcal P_3^{(i)}):\quad
\underset{\mathbf x}{\mathrm{max}}
\quad
\kappa^{(i)}
+
2\Re
\left\{
(\mathbf f^{(i)})^H\mathbf x
\right\}
-
\mathbf x^H
\mathbf Q^{(i)}
\mathbf x
\\
\mathrm{s.t.} \nonumber\\
&
\mathbf x^H\mathbf x
\le
\eta P_{\max},
\label{P3_power}
\\
&
\begin{aligned}[t]
\mathbf x^H
\mathbf Q_n^{(i)}
\mathbf x
-
2\Re
\left\{
(\mathbf f_n^{(i)})^H\mathbf x
\right\}
&\le
\kappa_n^{(i)}-R_n^{\min},
\:
\forall n\in\mathcal N,
\end{aligned}
\label{P3_qos}
\\
&
\begin{aligned}[t]
\mathbf x^H
\widehat{\mathbf H}_n
\mathbf x
&\le
2\left(
S_n^{(i)}+I_n^{(i)}
\right)
-\sigma^2,
\:
\forall n\in\mathcal N,
\end{aligned}
\label{P3_tr1}
\\
&
\begin{aligned}[t]
\mathbf x^H
\mathbf Q_n^{\mathrm{TR2},(i)}
\mathbf x
-
2\Re
\left\{
(\mathbf f_n^{\mathrm{TR2},(i)})^H
\mathbf x
\right\}
&\le
c_n^{\mathrm{TR2},(i)},
\:
\forall n\in\mathcal N.
\end{aligned}
\label{P3_tr2}
\end{align}
\end{subequations}

Problem~\eqref{P3} constitutes a structured QCQP in which the
objective function and all constraints admit a unified quadratic
representation. This structure forms the mathematical basis for the
consensus ADMM algorithm developed in the following subsection. For a fixed outer SCA iteration $i$, all surrogate-dependent
coefficients remain constant throughout the inner ADMM iterations.
Hence, the superscript $(i)$ is omitted from these coefficients in
the subsequent ADMM derivation for notational simplicity.

\subsection{Consensus Reformulation}

Although Problem~\eqref{P3} is convex, the quadratic objective and
multiple coupled constraints prevent a direct decomposition into
independent subproblems. To exploit the separable structure of the QCQP, local copies of the
optimization variable are introduced for each constraint, following
the consensus decomposition principle in~\cite{7517329}.

Specifically, let $\mathbf z_\ell$, $\ell\in\mathcal I$, denote
the local variable associated with the $\ell$th constraint, where
\begin{equation}
\begin{aligned}
&\mathcal I
=\\&\{\mathrm P,
\mathrm{QoS}_1,\ldots,\mathrm{QoS}_N,\mathrm{TR1}_1,\ldots,\mathrm{TR1}_N,\mathrm{TR2}_1,\ldots,\mathrm{TR2}_N\}.
\end{aligned}
\label{eq:indexset}
\end{equation}

Accordingly, $|\mathcal I|=1+3N$, and the consensus constraints
are imposed as
\begin{equation}
\mathbf z_\ell
=
\mathbf x,
\qquad
\forall \ell\in\mathcal I.
\label{eq:consensus}
\end{equation}

Using the above variable splitting,
Problem~\eqref{P3}
is equivalently reformulated as
\begin{subequations}
\label{P4}
\begin{align}
(\mathcal P_4):
\quad
&\underset{\mathbf x,\{\mathbf z_\ell\}}{\mathrm{maximize}}
\quad
\kappa
+
2\Re
\left\{
\mathbf f^H\mathbf x
\right\}
-
\mathbf x^H\mathbf Q\mathbf x
\\
\mathrm{s.t.}\quad
&
\mathbf z_{\mathrm P}^H
\mathbf z_{\mathrm P}
\le
\eta P_{\max},
\\
&
\mathbf z_\ell^H
\mathbf Q_\ell
\mathbf z_\ell
-
2\Re
\left\{
\mathbf f_\ell^H
\mathbf z_\ell
\right\}
\le
c_\ell,
\quad
\ell\in
\mathcal I\setminus\{\mathrm P\},
\\
&
\mathbf z_\ell
=
\mathbf x,
\qquad
\forall \ell\in\mathcal I.
\end{align}
\end{subequations}

Problem~\eqref{P4} is now in standard consensus form, where the
global variable appears only in the objective function, while each
constraint is handled independently through its corresponding local
variable.
\subsection{Augmented Lagrangian}

To apply ADMM to Problem~\eqref{P4}, we adopt the standard scaled
consensus formulation in~\cite{boyd2011distributed}. Problem~\eqref{P4} is
first converted into the equivalent convex minimization problem
\begin{subequations}
\label{PADMM}
\begin{align}
(\mathcal P_{\mathrm{ADMM}}):
&\underset{\mathbf x,\{\mathbf z_\ell\}}{\mathrm{minimize}}
\quad
\mathbf x^H\mathbf Q\mathbf x
-
2\Re
\left\{
\mathbf f^H\mathbf x
\right\}
-
\kappa
+
\sum_{\ell\in\mathcal I}
I_{\mathcal C_\ell}(\mathbf z_\ell)
\\
\mathrm{s.t.}\quad
&
\mathbf x
=
\mathbf z_\ell,
\qquad
\forall \ell\in\mathcal I.
\end{align}
\end{subequations}
where $I_{\mathcal C_\ell}(\cdot)$ denotes the indicator function of
the feasible set $\mathcal C_\ell$, defined as
\begin{equation}
I_{\mathcal C_\ell}(\mathbf z_\ell)
=
\left\{
\begin{array}{ll}
0,
&
\mathbf z_\ell\in\mathcal C_\ell,
\\
+\infty,
&
\text{otherwise}.
\end{array}
\right.
\label{eq:Indicator}
\end{equation}

The augmented Lagrangian associated with
Problem~\eqref{PADMM}
is given by
\begin{align}
\mathcal L_{\rho}
=
&
\mathbf x^H\mathbf Q\mathbf x
-
2\Re
\left\{
\mathbf f^H\mathbf x
\right\}
-
\kappa
+
\sum_{\ell\in\mathcal I}
I_{\mathcal C_\ell}(\mathbf z_\ell)
\nonumber\\
&
+
\sum_{\ell\in\mathcal I}
\Re
\left\{
\mathbf y_\ell^H
(\mathbf x-\mathbf z_\ell)
\right\}
+
\frac{\rho}{2}
\sum_{\ell\in\mathcal I}
\left\|
\mathbf x-\mathbf z_\ell
\right\|_2^2,
\label{eq:AL}
\end{align}
where $\mathbf y_\ell$ denotes the dual variable associated with the
$\ell$th consensus constraint.

Introducing the scaled dual variable
$\mathbf u_\ell=\mathbf y_\ell/\rho$, the augmented Lagrangian
can be equivalently rewritten as
\begin{align}
\mathcal L_{\rho}
=
&
\mathbf x^H\mathbf Q\mathbf x
-
2\Re
\left\{
\mathbf f^H\mathbf x
\right\}
-
\kappa
+
\sum_{\ell\in\mathcal I}
I_{\mathcal C_\ell}(\mathbf z_\ell)
\nonumber\\
&
+
\frac{\rho}{2}
\sum_{\ell\in\mathcal I}
\left\|
\mathbf x
-
\mathbf z_\ell
+
\mathbf u_\ell
\right\|_2^2
-
\frac{\rho}{2}
\sum_{\ell\in\mathcal I}
\left\|
\mathbf u_\ell
\right\|_2^2.
\label{eq:ScaledAL}
\end{align}

The scaled augmented Lagrangian in \eqref{eq:ScaledAL} serves as
the basis for deriving the proposed consensus ADMM iterations in
the following subsection.
\subsection{Global Variable Update}

At the $(k+1)$th ADMM iteration, the global beamforming variable is
obtained by minimizing the scaled augmented Lagrangian
\eqref{eq:ScaledAL} with respect to $\mathbf x$, while holding
$\{\mathbf z_\ell^{k},\mathbf u_\ell^{k}\}_{\ell\in\mathcal I}$
fixed, i.e.,
\begin{equation}
\mathbf x^{k+1}
=
\underset{\mathbf x}{\operatorname{arg\,min}}\;
\mathcal L_{\rho}
\left(
\mathbf x,
\{\mathbf z_\ell^{k}\}_{\ell\in\mathcal I},
\{\mathbf u_\ell^{k}\}_{\ell\in\mathcal I}
\right).
\label{eq:x_subproblem}
\end{equation}

Discarding all terms independent of $\mathbf{x}$, the resulting
subproblem is
\begin{equation}
\begin{aligned}
\mathbf{x}^{k+1}&=\\
{}
&\underset{\mathbf{x}}{\operatorname{arg\,min}}\;
\mathbf{x}^{H}\mathbf{Q}\mathbf{x}
-2\Re\!\left\{\mathbf{f}^{H}\mathbf{x}\right\} 
+
\frac{\rho}{2}
\sum_{\ell\in\mathcal{I}}
\left\|
\mathbf{x}-\mathbf{z}_{\ell}^{k}+\mathbf{u}_{\ell}^{k}
\right\|_{2}^{2}.
\end{aligned}
\label{eq:x_subproblem2}
\end{equation}

Expanding the quadratic penalty term gives
\begin{align}
\mathbf x^{k+1}
=
\underset{\mathbf x}{\operatorname{arg\,min}}
\quad&
\mathbf x^H
\left(
\mathbf Q
+
\frac{\rho|\mathcal I|}{2}\mathbf I
\right)
\mathbf x
\nonumber\\
&
-
2\Re
\left\{
\left(
\mathbf f
+
\frac{\rho}{2}
\sum_{\ell\in\mathcal I}
\left(
\mathbf z_\ell^k
-
\mathbf u_\ell^k
\right)
\right)^H
\mathbf x
\right\}.
\label{eq:x_subproblem3}
\end{align}
where the constant terms have been omitted since they do not
affect the minimizer.

Taking the Wirtinger derivative of
\eqref{eq:x_subproblem3}
with respect to
$\mathbf{x}^{*}$~\cite{5961645}
and equating the result to zero yields
\begin{equation}
\left(
\mathbf Q
+
\frac{\rho|\mathcal I|}{2}\mathbf I
\right)
\mathbf x^{k+1}
=
\mathbf f
+
\frac{\rho}{2}
\sum_{\ell\in\mathcal I}
\left(
\mathbf z_\ell^k
-
\mathbf u_\ell^k
\right).
\label{eq:x_stationarity}
\end{equation}

A direct unstructured solution of \eqref{eq:x_stationarity} requires
the factorization of an $MN\times MN$ linear system, which becomes
increasingly costly as the beamforming dimension grows. Fortunately,
the objective matrix admits a Kronecker product structure that
enables a significantly more efficient implementation.

Since each user-specific objective matrix is proportional to
$\widehat{\mathbf H}_n$, it can be written as
$\mathbf Q_n=\lambda_n\widehat{\mathbf H}_n$, where
\begin{equation}
\lambda_n
=
\frac{
S_n^{(i)}
}
{
I_n^{(i)}
\left(
S_n^{(i)}
+
I_n^{(i)}
\right)
}
+
cA_n^{(i)}
\beta_n^{(i)}.
\label{eq:lambda}
\end{equation}

Using
$\widehat{\mathbf H}_n=\mathbf I_N\otimes\mathbf H_n$,
the aggregate objective matrix can be written as
\begin{align}
\mathbf Q
&=
\sum_{n\in\mathcal N}
\lambda_n
(
\mathbf I_N
\otimes
\mathbf H_n
)
=
\mathbf I_N
\otimes
\left(
\sum_{n\in\mathcal N}
\lambda_n
\mathbf H_n
\right).
\label{eq:QKronecker}
\end{align}

Define
$\mathbf B=\sum_{n\in\mathcal N}\lambda_n\mathbf H_n$. Then, substituting \eqref{eq:QKronecker} into
\eqref{eq:x_stationarity}
yields
\begin{equation}
\left[
\mathbf I_N
\otimes
\left(
2\mathbf B
+
\rho |\mathcal I|
\mathbf I_M
\right)
\right]
\mathbf x^{k+1}
=
\mathbf r^{k},
\label{eq:KroneckerSystem}
\end{equation}
where
\begin{equation}
\mathbf r^k
=
2\mathbf f
+
\rho
\sum_{\ell\in\mathcal I}
\left(
\mathbf z_\ell^k
-
\mathbf u_\ell^k
\right).
\label{eq:rhs}
\end{equation}

Reshape the stacked beamforming vector
$\mathbf x^{k+1}$
into its matrix form
\begin{equation}
\mathbf X^{k+1}
=
\operatorname{unvec}
(
\mathbf x^{k+1}
)
\in
\mathbb C^{M\times N},
\label{eq:X}
\end{equation}

Similarly, reshape $\mathbf r^k$ into
$\mathbf R^k\in\mathbb C^{M\times N}$.
Using the identity
\begin{equation}
(
\mathbf I_N
\otimes
\mathbf A
)
\operatorname{vec}
(
\mathbf X
)
=
\operatorname{vec}
(
\mathbf A
\mathbf X
),
\label{eq:vecidentity}
\end{equation}

\eqref{eq:KroneckerSystem}
is equivalently expressed as
\begin{equation}
\left(
2\mathbf B
+
\rho |\mathcal I|
\mathbf I_M
\right)
\mathbf X^{k+1}
=
\mathbf R^{k}.
\label{eq:SmallLinearSystem}
\end{equation}

Therefore, the global beamforming update is obtained by solving a
common $M\times M$ linear system with $N$ right-hand sides,
\begin{equation}
\mathbf x_n^{k+1}
=
\left(
2\mathbf B
+
\rho |\mathcal I|
\mathbf I_M
\right)^{-1}
\mathbf r_n^{k},
\qquad
n=1,\ldots,N,
\label{eq:GlobalUpdate}
\end{equation}
where
$\mathbf r_n^{k}$
denotes the
$n$th
column of
$\mathbf R^{k}$. The inverse in \eqref{eq:GlobalUpdate} is not formed explicitly;
instead, the common $M\times M$ coefficient matrix is factorized
once, and the resulting factorization is reused across all $N$
beamforming blocks.
\subsection{Local Variable Update}

With the global beamforming variable fixed at
$\mathbf x^{k+1}$, each local variable is updated independently by
minimizing the scaled augmented Lagrangian with respect to
$\mathbf z_\ell$, i.e.,
\begin{equation}
\mathbf z_\ell^{k+1}
=
\underset{\mathbf z_\ell}{\operatorname{arg\,min}}\;
\mathcal L_\rho
\left(
\mathbf x^{k+1},
\mathbf z_\ell,
\mathbf u_\ell^k
\right),
\qquad
\forall \ell\in\mathcal I.
\label{eq:z_problem}
\end{equation}

Since the local variables are completely decoupled, the resulting
projection subproblems can be solved independently. The
transmit-power constraint admits a Euclidean projection, whereas
the QoS, TR1 and TR2 constraints share an identical QCQP
structure and are therefore handled by a unified projection
operator.

\subsubsection{Power Projection}

The power projection is obtained by solving
\begin{subequations}
\label{P5}
\begin{align}
\mathbf z_{\mathrm P}^{k+1}
=\quad
&\underset{\mathbf z}{\operatorname{arg\,min}}
\quad
\frac{1}{2}
\left\|
\mathbf z-\mathbf v_{\mathrm P}
\right\|_2^2
\\
\mathrm{s.t.}\quad&
\|\mathbf z\|_2^2
\le
\eta P_{\max}.
\end{align}
\end{subequations}
where
\begin{equation}
\mathbf v_{\mathrm P}
=
\mathbf x^{k+1}
+
\mathbf u_{\mathrm P}^{k}.
\label{eq:vPower}
\end{equation}

Problem~\eqref{P5} is the Euclidean projection onto the feasible
power ball and therefore admits the closed-form solution
\begin{equation}
\mathbf z_{\mathrm P}^{k+1}
=
\left\{
\begin{array}{ll}
\mathbf v_{\mathrm P},
&
\|\mathbf v_{\mathrm P}\|_2
\le
\sqrt{\eta P_{\max}},
\\[2mm]
\displaystyle
\frac{\sqrt{\eta P_{\max}}}
{\|\mathbf v_{\mathrm P}\|_2}
\mathbf v_{\mathrm P},
&
\text{otherwise}.
\end{array}
\right.
\label{eq:PowerProjection}
\end{equation}

\subsubsection{Unified QCQP Projection}

The remaining local variables correspond to the QoS, TR1 and
TR2 constraints, all of which can be expressed in the unified
QCQP form
\begin{subequations}
\label{P6}
\begin{align}
\mathbf z^{k+1}
=\quad
&\underset{\mathbf z}{\operatorname{arg\,min}}
\quad
\frac{1}{2}
\left\|
\mathbf z-\mathbf v
\right\|_2^2
\\
\mathrm{s.t.}\quad&
\mathbf z^H
\mathbf Q
\mathbf z
-
2\Re
\left\{
\mathbf f^H
\mathbf z
\right\}
\le
c.
\end{align}
\end{subequations}

Introducing the Lagrange multiplier
$\mu\ge0$,
the corresponding Lagrangian is
\begin{align}
\mathcal L
=
&
\frac12
\|
\mathbf z
-
\mathbf v
\|_2^2
+
\mu
\left(
\mathbf z^H
\mathbf Q
\mathbf z
-
2
\Re
\left\{
\mathbf f^H
\mathbf z
\right\}
-
c
\right).
\label{eq:ProjectionLagrangian}
\end{align}

Taking the Wirtinger derivative with respect to
$\mathbf z^{*}$
and setting the result to zero gives
\begin{equation}
\left(
\mathbf I
+
2\mu
\mathbf Q
\right)
\mathbf z
=
\mathbf v
+
2\mu
\mathbf f.
\label{eq:ProjectionStationarity}
\end{equation}

Accordingly, the projected point is
\begin{equation}
\mathbf z
=
\left(
\mathbf I
+
2\mu
\mathbf Q
\right)^{-1}
\left(
\mathbf v
+
2\mu
\mathbf f
\right).
\label{eq:ProjectionSolution}
\end{equation}

Substituting \eqref{eq:ProjectionSolution} into the active
constraint yields a scalar nonlinear equation in $\mu$. The
corresponding multiplier is efficiently obtained using Newton's
method~\cite{nocedal2006numerical}. If the unconstrained point is feasible, the projection is obtained
directly as $\mathbf z=\mathbf v$ with $\mu=0$. Otherwise, the
scalar multiplier is computed using Newton's method initialized at
$\mu=0.1$, with nonnegativity enforced at each iteration. The
iteration terminates when the absolute constraint residual falls
below $10^{-10}$, with a maximum of 500 iterations.

Furthermore, the QCQP matrices satisfy
\begin{equation}
\mathbf Q
=
\lambda
\left(
\mathbf I_N
\otimes
\mathbf H
\right),
\label{eq:ProjectionQ}
\end{equation}
where the scalar
$\lambda$
depends on the corresponding constraint.
Consequently,
\begin{equation}
\mathbf I
+
2\mu
\mathbf Q
=
\mathbf I_N
\otimes
\left(
\mathbf I_M
+
2\mu\lambda
\mathbf H
\right).
\label{eq:ProjectionKronecker}
\end{equation}

Since
$\mathbf H=\mathbf h\mathbf h^H$
is rank one, the inverse of
$\mathbf I_M+2\mu\lambda\mathbf H$
is evaluated using the Sherman--Morrison formula \cite{golub2013matrix},
\begin{equation}
\left(
\mathbf I_M
+
2\mu\lambda
\mathbf H
\right)^{-1}
=
\mathbf I_M
-
\frac{
2\mu\lambda
}
{
1
+
2\mu\lambda
\|\mathbf h\|_2^2
}
\mathbf H.
\label{eq:Sherman}
\end{equation}

Using
\eqref{eq:ProjectionKronecker}
and
\eqref{eq:Sherman},
the inverse action is evaluated directly
through rank-one inner products and scaled vector operations across
the $N$ beamforming blocks, without explicitly forming the
$M\times M$ inverse. Consequently,
the inversion of the full
$MN\times MN$
matrix is completely avoided, resulting in a significantly lower
computational complexity. It is worth noting that the QoS, TR1 and TR2 constraints differ
only in the corresponding triplet
$(\mathbf Q,\mathbf f,c)$.
Consequently, all three constraints are handled by the same QCQP
projection routine, thereby eliminating the need for separate
projection algorithms.

\subsection{Dual Variable Update}

After updating the global and local variables, the scaled dual
variables are updated according to the consensus residuals. The
dual update associated with the $\ell$th constraint is
\begin{equation}
\mathbf u_\ell^{k+1}
=
\mathbf u_\ell^k
+
\mathbf x^{k+1}
-
\mathbf z_\ell^{k+1},
\qquad
\forall \ell\in\mathcal I.
\label{eq:DualUpdate}
\end{equation}

The primal residual is defined as
\begin{equation}
\mathbf r_{\mathrm p}^{k+1}
=
\left\{
\mathbf x^{k+1}
-
\mathbf z_\ell^{k+1}
\right\}_{\ell\in\mathcal I}.
\label{eq:PrimalResidual}
\end{equation}
whereas the dual residual measure, based on successive changes in the local consensus variables, is given as

\begin{equation}
\mathbf r_{\mathrm d}^{k+1}
=
\rho
\left\{
\mathbf z_\ell^{k+1}
-
\mathbf z_\ell^k
\right\}_{\ell\in\mathcal I}.
\label{eq:DualResidual}
\end{equation}

The proposed algorithm terminates when both residuals satisfy the
prescribed convergence tolerance. The penalty parameter is adaptively updated using the residual-balancing
strategy in~\cite{boyd2011distributed}. Specifically,
\begin{equation}
\rho^{k+1}
=
\begin{cases}
2\rho^k, &
\|\mathbf r_{\mathrm p}^{k}\|_2
>
10\|\mathbf r_{\mathrm d}^{k}\|_2,
\\
\rho^k/2, &
\|\mathbf r_{\mathrm d}^{k}\|_2
>
10\|\mathbf r_{\mathrm p}^{k}\|_2,
\\
\rho^k, & \text{otherwise}.
\end{cases}
\label{eq:rho_update}
\end{equation}
The update in \eqref{eq:rho_update} is evaluated every 20 ADMM
iterations, and the scaled dual variables are rescaled whenever
$\rho$ is modified. At each SCA iteration, Problem~\eqref{P4} is a convex consensus
problem equivalent to Problem~\eqref{P3}. The proposed inner
iterations therefore follow the standard consensus-ADMM framework
for convex optimization~\cite{boyd2011distributed}, with practical
termination determined by the prescribed primal and dual residual
tolerances. The overall structured consensus ADMM-SCA procedure is
summarized in Algorithm~\ref{alg:ProposedADMM}.
\begin{table*}[!t]
\centering
\caption{Computational complexity for solving one convex QCQP.}
\label{tab:complexity}
\footnotesize
\setlength{\tabcolsep}{4pt}
\renewcommand{\arraystretch}{1.18}

\begin{tabularx}{\textwidth}{
@{}
>{\raggedright\arraybackslash}p{3.2cm}
>{\raggedright\arraybackslash}p{4.4cm}
>{\centering\arraybackslash}X
@{}
}
\toprule
\textbf{Method}
&
\textbf{Dominant Operations}
&
\textbf{Complexity per QCQP Solve}
\\
\midrule

Standard ADMM
&
Dense $MN\times MN$ global solve and
$3N$ full-dimensional QCQP projections
&
$\mathcal{O}\!\left(
N_{\mathrm{ADMM}}^{\mathrm{std}}
(1+3NI_{\mu})(MN)^3
\right)$
\\

Proposed structured ADMM
&
Shared $M\times M$ factorization and
rank-one local QCQP projections
&
$\mathcal{O}\!\left(
N_{\mathrm{ADMM}}^{\mathrm{prop}}
\left[
M^3+NM^2+3I_\mu MN^2
\right]
\right)$
\\

CVX/MOSEK
(interior-point method)
&
Interior-point Newton/KKT system solution
&
$\mathcal{O}\!\left(
\sqrt{3N+1}
\left[
(MN)^3
+(3N+1)(MN)^2
+(3N+1)^2(MN)
\right]
\right)$
\\

\bottomrule
\end{tabularx}
\end{table*}

\begin{algorithm}[!t]
\caption{Proposed Consensus ADMM Beamforming}
\label{alg:ProposedADMM}
\begin{algorithmic}[1]

\Require Initial SCA expansion point
$\{\mathbf w_n^{(0)}\}_{n\in\mathcal N}$,
initial penalty parameter $\rho$,
maximum numbers of SCA and ADMM iterations
$I_{\max}$ and $K_{\max}$,
and ADMM residual tolerances.

\Ensure Optimized beamforming vectors
$\{\mathbf w_n^\star\}_{n\in\mathcal N}$.

\For{$i=0,\ldots,I_{\max}-1$}

    \State Construct the surrogate objective and constraint parameters for the current SCA expansion point.

    \State Initialize
$\mathbf{x}^{0}$,
$\{\mathbf z_\ell^{0}\}_{\ell\in\mathcal I}$,
and
$\{\mathbf u_\ell^{0}\}_{\ell\in\mathcal I}$
from the current SCA expansion point.

    \For{$k=0,\ldots,K_{\max}-1$}

        \State Update the global variable using \eqref{eq:GlobalUpdate}.

        \State Update the local variables using
        \eqref{eq:PowerProjection}
        and
        \eqref{eq:ProjectionSolution}.

        \State Update the scaled dual variables using
        \eqref{eq:DualUpdate}.

        \State Evaluate the primal and dual residuals.

        \If{the convergence criterion is satisfied}
            \State \textbf{break}
        \EndIf

        \State Update $\rho$ according to \eqref{eq:rho_update} and
rescale the scaled dual variables, if necessary.

    \EndFor

    \State Update the SCA expansion point using the ADMM solution.

    \If{$\left|
R_{\mathrm{sum}}^{(i)}
-
R_{\mathrm{sum}}^{(i-1)}
\right|<10^{-3}$}
        \State \textbf{break}
    \EndIf

\EndFor

\State \Return
$\{\mathbf w_n^\star\}_{n\in\mathcal N}$.

\end{algorithmic}
\end{algorithm}

\subsection{Computational Complexity}

Table~\ref{tab:complexity} summarizes the computational complexity
of solving one convex QCQP subproblem using standard ADMM, the
proposed structured ADMM, and CVX/MOSEK. Let
$N_{\mathrm{ADMM}}^{\mathrm{std}}$ and
$N_{\mathrm{ADMM}}^{\mathrm{prop}}$ denote the numbers of inner
iterations required by the standard and proposed ADMM
implementations, respectively. Moreover, $I_\mu$ denotes the number
of Newton iterations required to determine the scalar multiplier in
an active local QCQP projection.

For standard ADMM, the global update requires solving a dense
$MN\times MN$ linear system, while each active user-dependent
projection involves full-dimensional linear-system solves during
the Newton iterations. This results in a per-iteration complexity of
$\mathcal{O}((1+3NI_\mu)(MN)^3)$.

In the proposed implementation, the Kronecker structure reduces the
global update to a shared $M\times M$ factorization, which is reused
across the $N$ beamforming blocks, resulting in
$\mathcal{O}(M^3+NM^2)$ complexity. Moreover, the rank-one channel
structure enables the Sherman--Morrison inverse action to be
evaluated directly through inner products and scaled vector
operations. The $3N$ local projections therefore require
$O(3I_\mu MN^2)$ operations per ADMM iteration. Accounting
for the corresponding numbers of inner iterations yields the
standard and proposed ADMM complexities reported in
Table~\ref{tab:complexity}.

For CVX/MOSEK, Table~\ref{tab:complexity} reports the generic dense
interior-point complexity estimate for Problem~\eqref{P3}, with
optimization dimension $MN$ and $3N+1$ quadratic constraints
~\cite{vandenberghe2004convex}. Thus, the proposed implementation
avoids full-dimensional factorizations in both the global and local
ADMM updates. In addition, the local projections are mutually
independent and can be executed in parallel.

\section{Numerical Results}
\label{sec:numerical_results}

\begin{figure*}[t]
\centering
\begin{subfigure}[t]{0.49\textwidth}
    \centering
    \includegraphics[width=\linewidth]{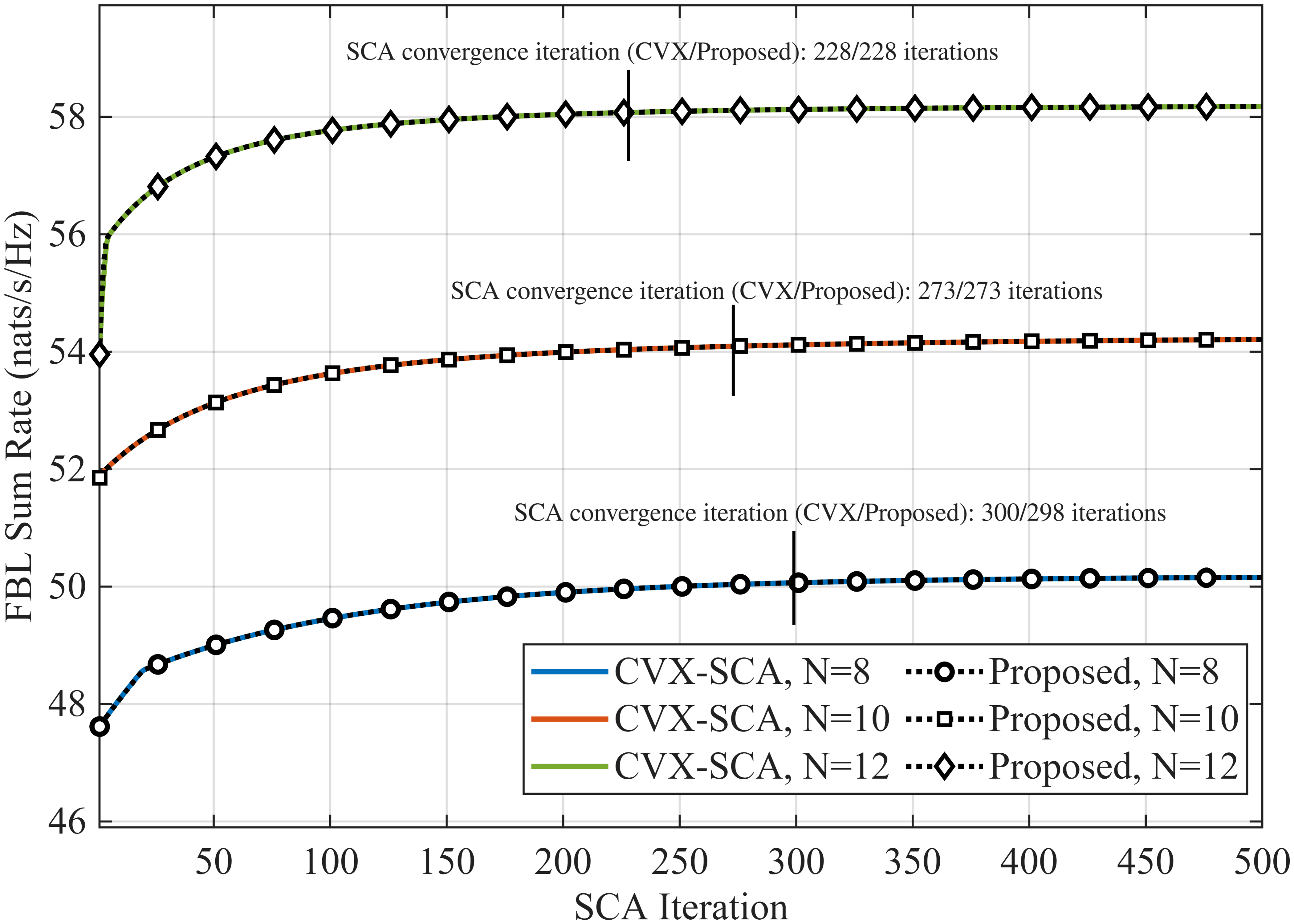}
    \caption{Varying the number of users with $M=16$.}
    \label{fig:sca_N}
\end{subfigure}
\hfill
\begin{subfigure}[t]{0.48\textwidth}
    \centering
    \includegraphics[width=\linewidth]{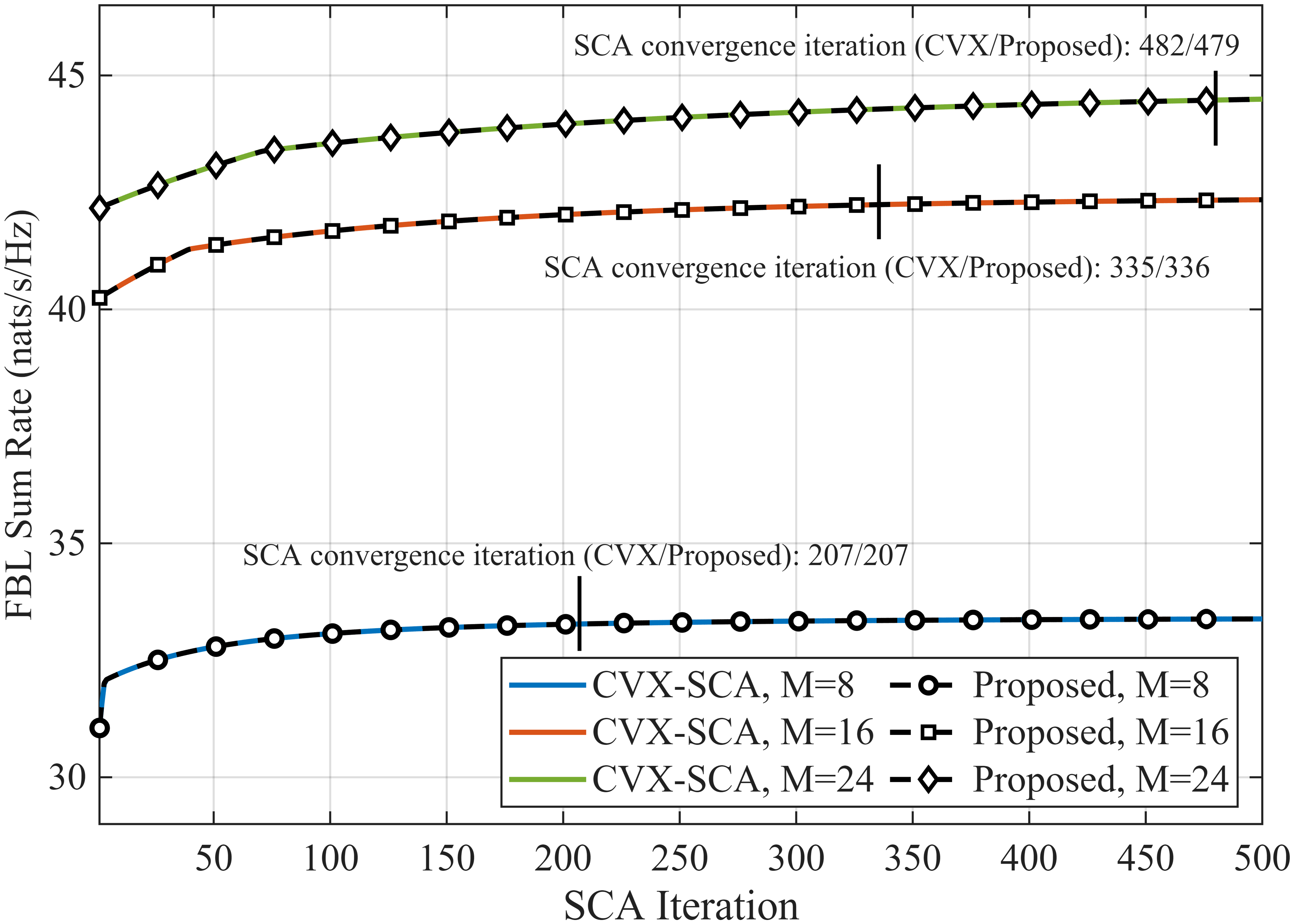}
    \caption{Varying the number of BS antennas with $N=6$.}
    \label{fig:sca_M}
\end{subfigure}
\caption{FBL sum-rate convergence of CVX-SCA and the proposed
structured ADMM-SCA for different system dimensions.}
\label{fig:sca_convergence}
\end{figure*}

This section evaluates the solution accuracy, convergence behavior,
and computational efficiency of the proposed structured consensus
ADMM-SCA framework. Two reference implementations are considered.
CVX-SCA solves each convex SCA subproblem using CVX/MOSEK and
serves as the general-purpose numerical benchmark, whereas standard
ADMM-SCA applies the same consensus formulation to the same QCQP
but evaluates the global and local updates directly in the full
$MN$-dimensional representation, without exploiting the Kronecker
and rank-one channel structures. The latter comparison
isolates the computational gain attributable to the proposed
structural exploitation rather than to the use of ADMM itself. For each comparison, all methods use the same channel realization, initialization, and
outer-SCA stopping criterion.

We consider a downlink cellular IoT network in which
$N$ single-antenna URLLC IoT nodes are independently
and uniformly distributed over an annular cell
with inner and outer radii of 50~m and 200~m,
respectively. The channels follow independent Rayleigh fading with the
distance-dependent large-scale path-loss model adopted
in~\cite{9246274},
\begin{equation}
\beta_n[\mathrm{dB}]
=
-35.3-10\alpha\log_{10}(d_n),
\end{equation}
where $d_n$ denotes the BS--user distance and $\alpha=3.76$. Unless
otherwise stated, the BS is equipped with $M=16$ antennas. The
system bandwidth is $1$~MHz, the noise power spectral density is
$-174$~dBm/Hz, the blocklength is $L=1000$ channel uses, and the
target decoding error probability is $\varepsilon=10^{-5}$. The
minimum FBL rate is $R_n^{\min}=0.1$~nats/s/Hz for all users. The maximum RF transmit power is $24$~dBm and the power-amplifier
efficiency is $\eta=0.5$. Accordingly, the maximum DC power budget
is $P_{\max}=P_{\mathrm{RF,max}}/\eta$. The initial ADMM penalty parameter is set to $\rho=128$, selected
empirically from a preliminary parameter sweep, and is subsequently
adapted according to \eqref{eq:rho_update}. The primal and dual residual tolerances are set to $10^{-4}$. All simulations are performed in MATLAB R2025b on a Windows 11
workstation equipped with an Intel Core i9-14900KF processor and
64~GB of RAM. CVX 2.2 with MOSEK 9.1.9 is used for the CVX-SCA
benchmark. Each convex SCA subproblem is solved by the inner ADMM iterations
until the prescribed primal and dual residual tolerances are
satisfied. The resulting beamforming solution is then used as the
expansion point for the subsequent SCA iteration. The outer-SCA
convergence iteration is defined as the first iteration $i$ satisfying
$\left|R_{\mathrm{sum}}^{(i)}-R_{\mathrm{sum}}^{(i-1)}\right|
<10^{-3}$, where $R_{\mathrm{sum}}^{(i)}$ denotes the FBL sum rate.
For Fig.~\ref{fig:sca_convergence}, the trajectories are evaluated
over 500 SCA iterations, and the marked points indicate the
corresponding convergence iterations.

We first examine whether the proposed structured ADMM solver
preserves the convergence behavior of the CVX/MOSEK-based SCA
implementation. Fig.~\ref{fig:sca_convergence} shows the FBL sum
rate versus the SCA iteration when varying the number of users and
BS antennas. For $M=16$, Fig.~\ref{fig:sca_N} shows that the
proposed ADMM-SCA method closely follows CVX-SCA for
$N\in\{8,10,12\}$ and attains essentially the same FBL sum rate.
For $N=8$, the convergence iterations are 300 and 298 for CVX-SCA
and the proposed method, respectively. The
corresponding iteration counts are 273/273 for $N=10$ and 228/228
for $N=12$. Thus, replacing CVX/MOSEK with the proposed structured
ADMM solver does not introduce an appreciable change in the number
of outer SCA iterations required for convergence. A similar agreement is observed when the number of BS antennas is
varied with $N=6$. As shown in Fig.~\ref{fig:sca_M}, the
CVX-SCA/proposed ADMM-SCA convergence iteration counts are 207/207,
335/336, and 482/479 for $M=8$, $16$, and $24$, respectively.
The two methods also attain essentially identical final FBL sum
rates. These results show that the proposed structured implementation
preserves the attained solution accuracy and exhibits essentially
the same outer-SCA convergence behavior as the CVX/MOSEK benchmark
for the considered system configurations. Consequently,
any computational reduction observed in the following results is
not obtained at the expense of additional SCA iterations or a
degradation in the attained beamforming solution.
\begin{figure*}[t]
\centering

\begin{minipage}[t]{0.48\textwidth}
    \centering
    \includegraphics[width=\linewidth]{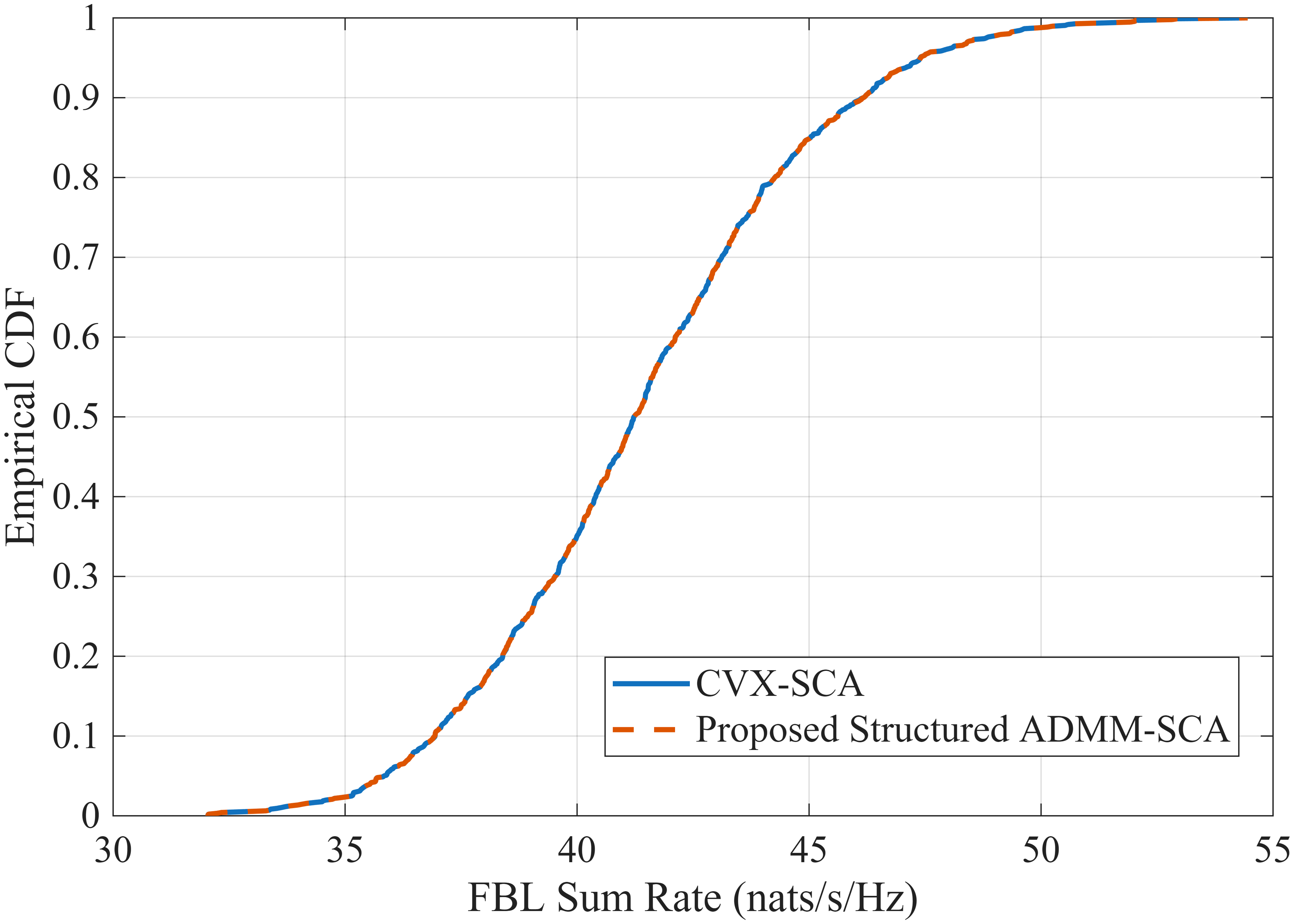}
    \caption{Empirical CDF of the achieved FBL sum rate over $1000$
    independent channel realizations for $M=16$ and $N=8$.}
    \label{fig:cdf_rate}
\end{minipage}
\hfill
\begin{minipage}[t]{0.48\textwidth}
    \centering
    \includegraphics[width=\linewidth]{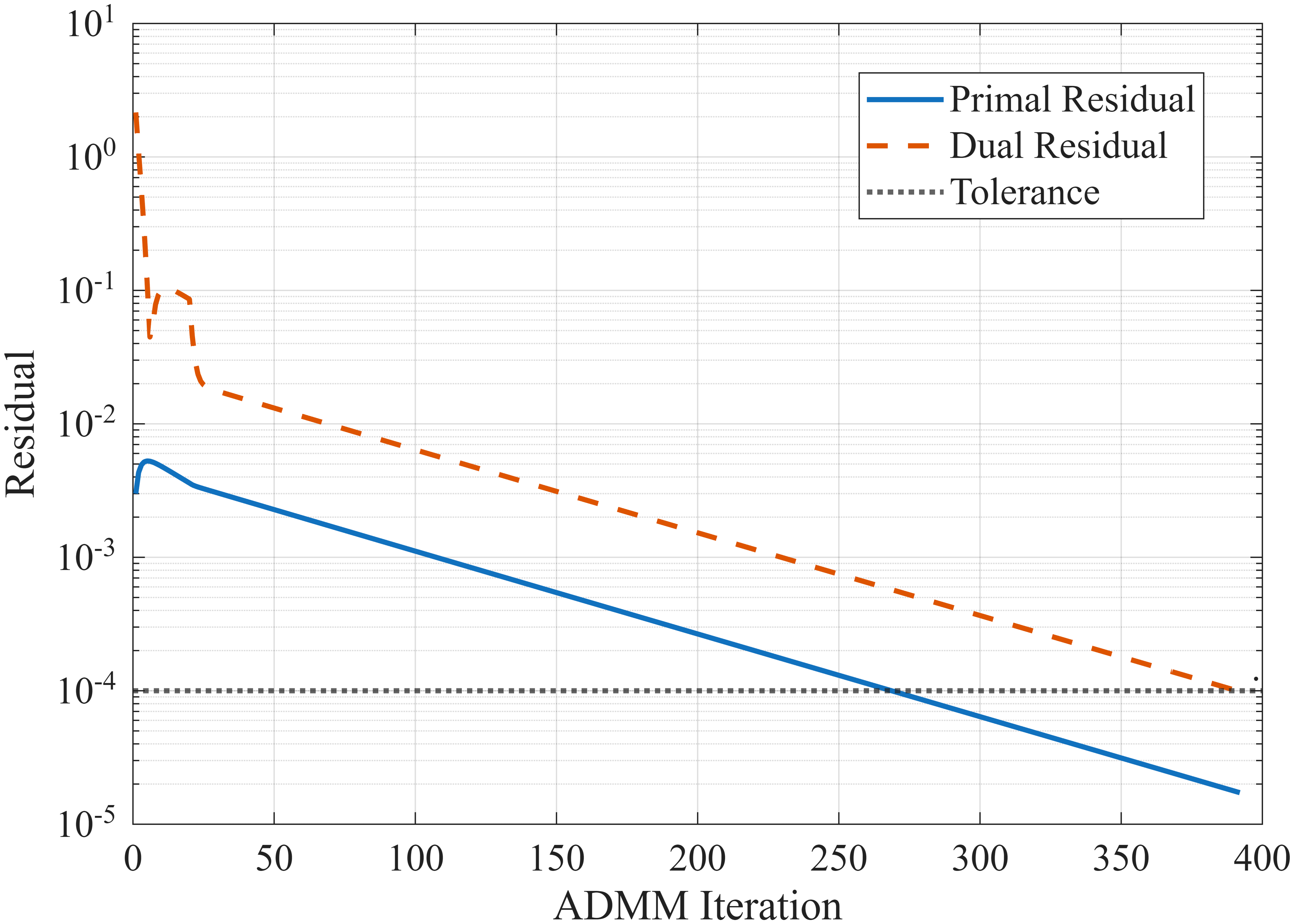}
    \caption{Primal and dual residual convergence of the proposed
    structured ADMM for a representative convex SCA subproblem with
    $M=16$ and $N=8$.}
    \label{fig:admm_residual}
\end{minipage}

\end{figure*}

To assess whether the agreement between CVX-SCA and the proposed
structured ADMM-SCA persists across different channel conditions,
Fig.~\ref{fig:cdf_rate} shows the empirical CDF of the final FBL
sum rate for $M=16$ and $N=8$ over $1000$ independent channel
realizations. For each realization, both methods use the same
channel and initialization. The two empirical CDFs are nearly
indistinguishable over the entire observed range, demonstrating
that the close agreement in the attained FBL sum rate is maintained
across independent channel realizations. This statistical result
complements the representative convergence trajectories in
Fig.~\ref{fig:sca_convergence}.

To examine the convergence of the inner solver,
Fig.~\ref{fig:admm_residual} shows the primal and dual residuals of
the proposed structured ADMM for a representative convex SCA
subproblem with $M=16$ and $N=8$. Both residuals decrease with the
ADMM iterations and eventually fall below the prescribed tolerance
of $10^{-4}$. The primal residual reaches the tolerance earlier,
while the iterations continue until the dual residual also satisfies
the stopping criterion. The decay of both residuals below the
prescribed tolerance confirms the numerical convergence of the inner
ADMM iterations for the considered subproblem and complements the
outer-SCA convergence results in Fig.~\ref{fig:sca_convergence}.

Table~\ref{tab:solution_accuracy} further compares the attained
solution accuracy and feasibility of the three implementations for
$M=16$ and $N=8$. CVX-SCA achieves an FBL sum rate of
$49.850$~nats/s/Hz, whereas standard ADMM-SCA and the proposed
structured ADMM-SCA both achieve $49.848$~nats/s/Hz, corresponding
to a relative difference of approximately $0.004\%$. All three methods satisfy the QoS constraints. The ADMM-based
implementations exhibit a power-constraint violation of only
$7.91\times10^{-5}$~W, which is numerically negligible relative
to the RF transmit-power budget $\eta P_{\max}$. This small residual feasibility error is
consistent with terminating the inner ADMM iterations at finite
primal and dual residual tolerances.
Moreover, standard and structured ADMM attain identical objective
values and feasibility levels, confirming that the proposed
structural exploitation reduces the computational cost of the
consensus updates without introducing an additional loss in solution
accuracy.
\begin{table}[t]
\centering
\caption{Solution Accuracy and Feasibility Comparison ($M=16$, $N=8$)}
\label{tab:solution_accuracy}
\footnotesize
\setlength{\tabcolsep}{2pt}
\renewcommand{\arraystretch}{1.15}
\begin{tabularx}{\columnwidth}{@{}>{\raggedright\arraybackslash}Xccc@{}}
\hline
\textbf{Method} &
\textbf{\begin{tabular}[c]{@{}c@{}}FBL Sum Rate\\(nats/s/Hz)\end{tabular}} &
\textbf{\begin{tabular}[c]{@{}c@{}}QoS\\Violation\end{tabular}} &
\textbf{\begin{tabular}[c]{@{}c@{}}Power Violation\\(W)\end{tabular}} \\
\hline
CVX-SCA
& 49.850 & $0$ & $0$ \\

Standard ADMM-SCA
& 49.848 & $0$ & $7.91\times10^{-5}$ \\

Proposed Structured ADMM-SCA
& 49.848 & $0$ & $7.91\times10^{-5}$ \\
\hline
\end{tabularx}
\end{table}

\begin{figure*}[t]
\centering

\begin{subfigure}[t]{0.48\textwidth}
    \centering
    \includegraphics[width=\linewidth]
    {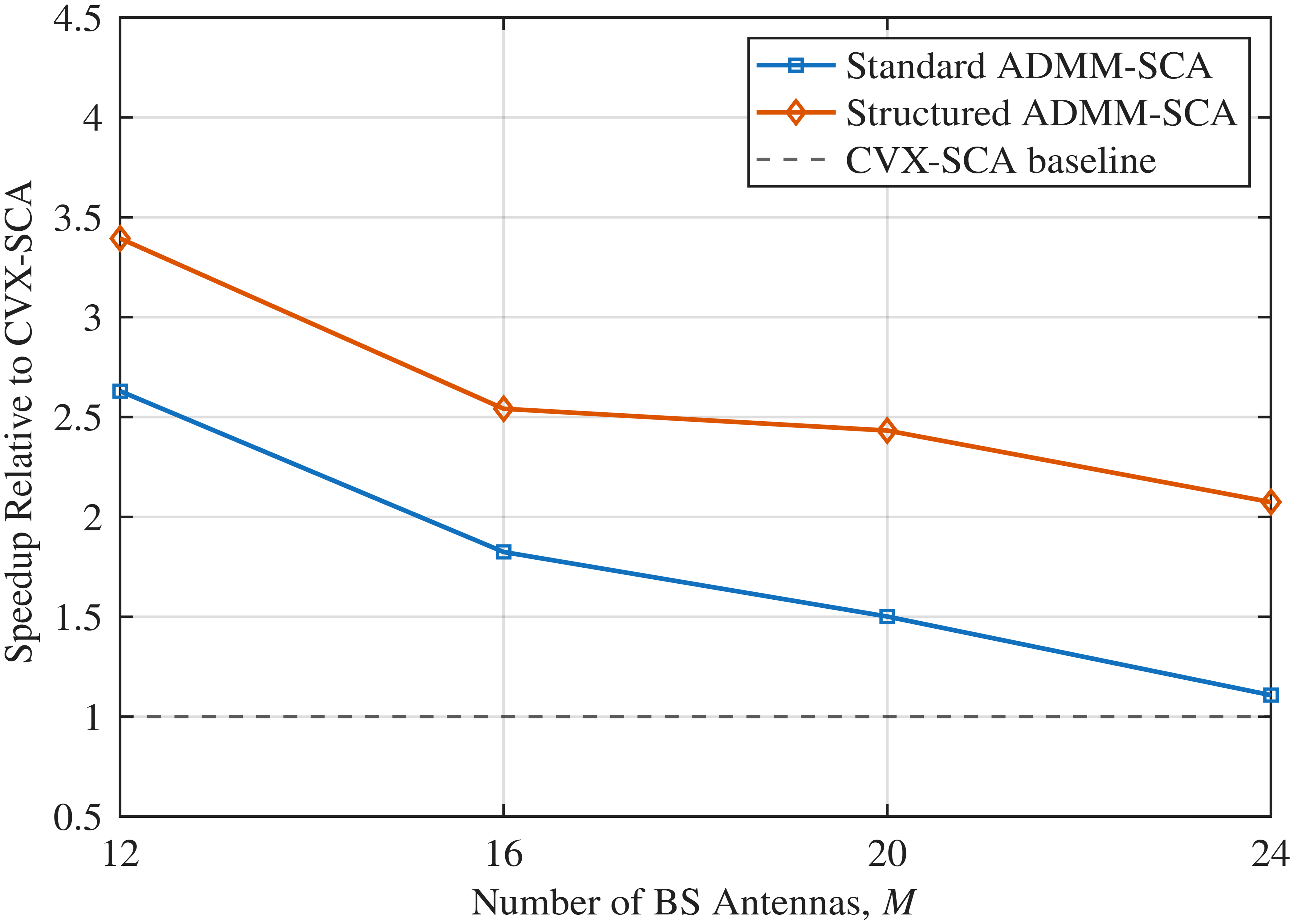}
    \caption{Speedup versus $M$ with $N=6$.}
    \label{fig:cvx_speedup_M}
\end{subfigure}
\hfill
\begin{subfigure}[t]{0.48\textwidth}
    \centering
    \includegraphics[width=\linewidth]
    {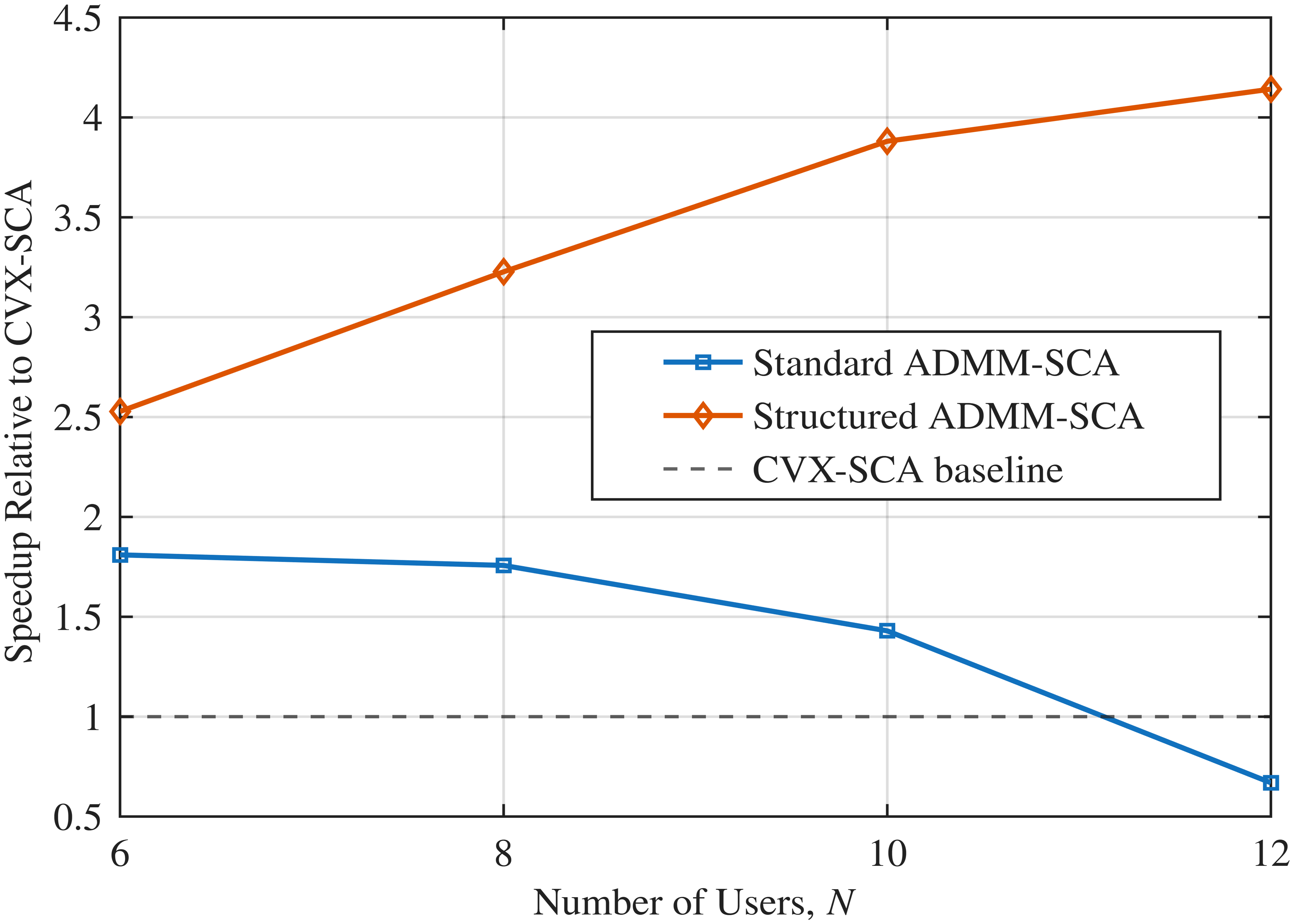}
    \caption{Speedup versus $N$ with $M=16$.}
    \label{fig:cvx_speedup_N}
\end{subfigure}

\caption{End-to-end computational speedup of standard ADMM-SCA
and the proposed structured ADMM-SCA relative to CVX-SCA under
antenna and user scaling. The $1\times$ reference corresponds to
CVX-SCA.}
\label{fig:cvx_speedup}

\end{figure*}

\begin{figure*}[t]
\centering
\begin{subfigure}[t]{0.48\textwidth}
    \centering
    \includegraphics[width=\linewidth]{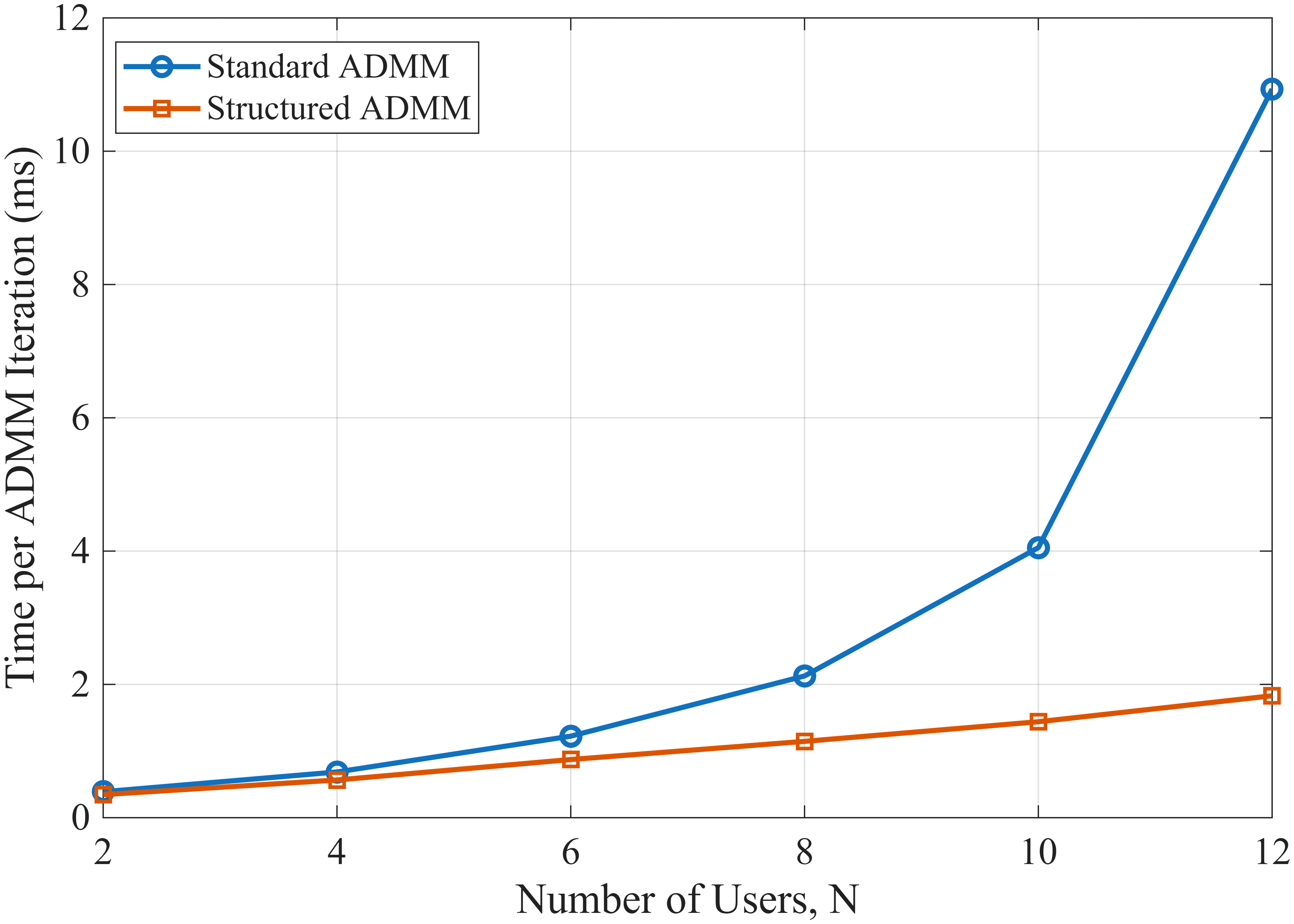}
    \caption{Per-iteration cost versus $N$ with $M=16$.}
    \label{fig:admm_cost_N}
\end{subfigure}
\hfill
\begin{subfigure}[t]{0.48\textwidth}
    \centering
    \includegraphics[width=\linewidth]{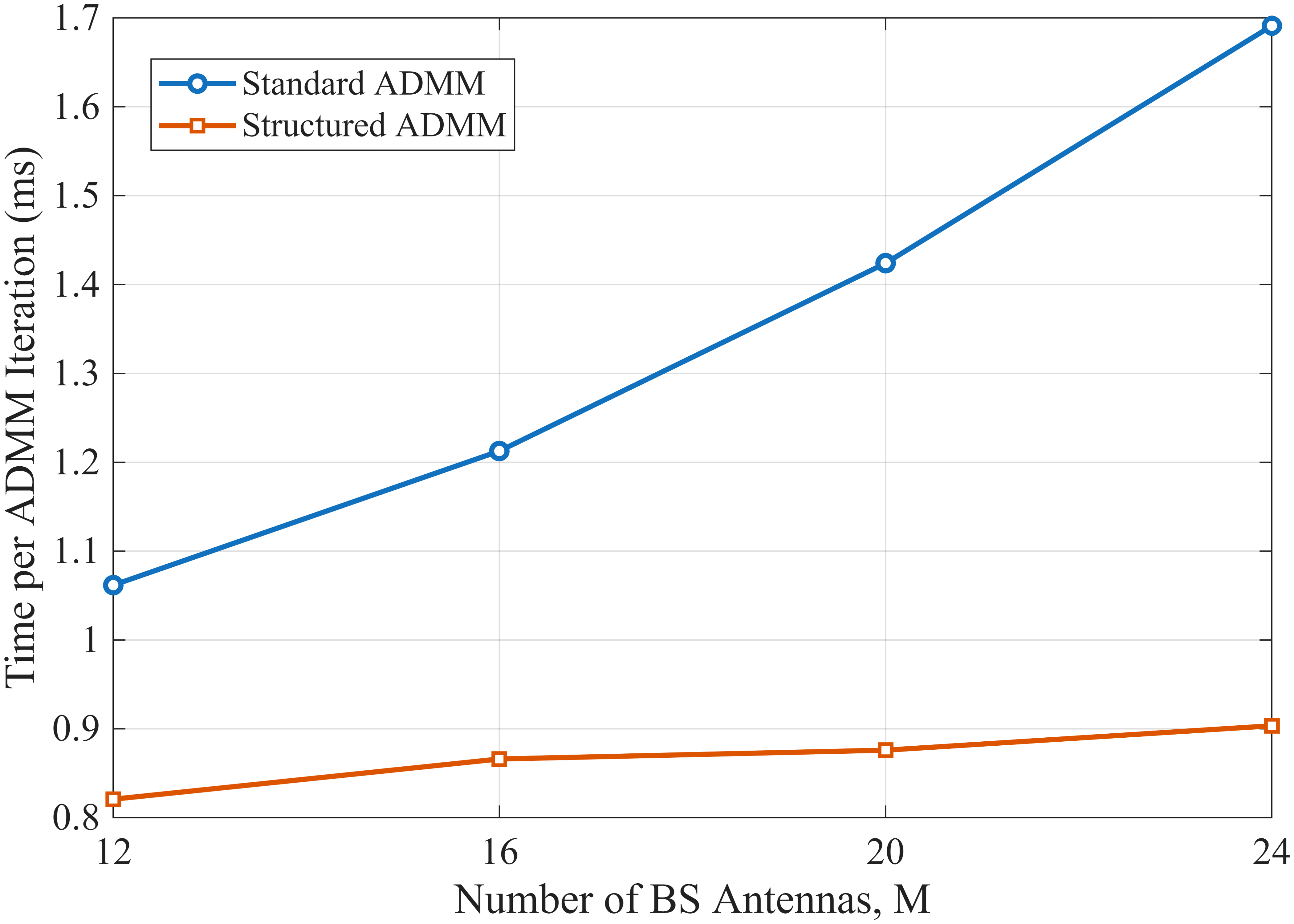}
    \caption{Per-iteration cost versus $M$ with $N=6$.}
    \label{fig:admm_cost_M}
\end{subfigure}
\caption{Average computational cost per ADMM iteration for standard
and structured ADMM.}
\label{fig:admm_cost}
\end{figure*} 

Having verified the close agreement of the proposed structured
ADMM-SCA with CVX-SCA in solution accuracy and outer-SCA convergence
for the considered configurations, we next evaluate its end-to-end
computational efficiency.
Fig.~\ref{fig:cvx_speedup} reports the computational speedup relative
to CVX-SCA, defined as
$T_{\mathrm{CVX}}/T_{\mathrm{method}}$, where values above unity
indicate a computational advantage over CVX-SCA. The reported results include the SCA iterations required to satisfy the outer-SCA stopping criterion and are averaged over independent Monte Carlo channel realizations. For fixed $N=6$, Fig.~\ref{fig:cvx_speedup_M} shows that the
proposed structured ADMM-SCA achieves a speedup of approximately
$3.4\times$ at $M=12$ and remains above $2\times$ at $M=24$.
The reduction in speedup with increasing $M$ is consistent with the
antenna dependence of the structured updates. Although the Kronecker decomposition reduces the global linear
system from dimension $MN$ to $M$, the shared $M\times M$
factorization still incurs an $\mathcal{O}(M^3)$ cost, while the
$N$ associated block solves scale as $\mathcal{O}(NM^2)$. Each rank-one local projection, by contrast,
requires $\mathcal O(I_\mu NM)$ operations. Standard
ADMM-SCA exhibits a more pronounced degradation because its updates
operate on the full $MN$-dimensional representation, causing its
speedup to approach the CVX-SCA reference as $M$ increases.

A different scaling behavior is observed in
Fig.~\ref{fig:cvx_speedup_N}. With $M=16$ fixed, the speedup of the
proposed structured ADMM-SCA increases from approximately
$2.5\times$ at $N=6$ to more than $4\times$ at $N=12$, whereas
that of standard ADMM-SCA decreases with $N$ and falls below unity
at $N=12$. With fixed $M$, increasing $N$ does not change the
dimension of the $M\times M$ matrix factorized in the structured
global update, and the same factorization is reused across all
$N$ beamforming blocks. The additional computational cost therefore
arises primarily from the block solves and user-dependent local
projections. In contrast, standard ADMM-SCA retains operations on
the full $MN$-dimensional representation, while the number of local
constraint updates simultaneously increases as
$|\mathcal I|=1+3N$. The results therefore distinguish the benefit of the proposed
structural exploitation from that of applying ADMM alone. Standard
ADMM-SCA progressively loses its computational advantage over
CVX-SCA as $N$ increases, whereas the proposed structured
implementation exhibits an increasing speedup. This behavior is
consistent with the complexity analysis and demonstrates the
computational benefit of the Kronecker-structured global update and
rank-one local projections.

\begin{figure*}[t]
\centering
\begin{subfigure}[t]{0.48\textwidth}
    \centering
    \includegraphics[width=\linewidth]{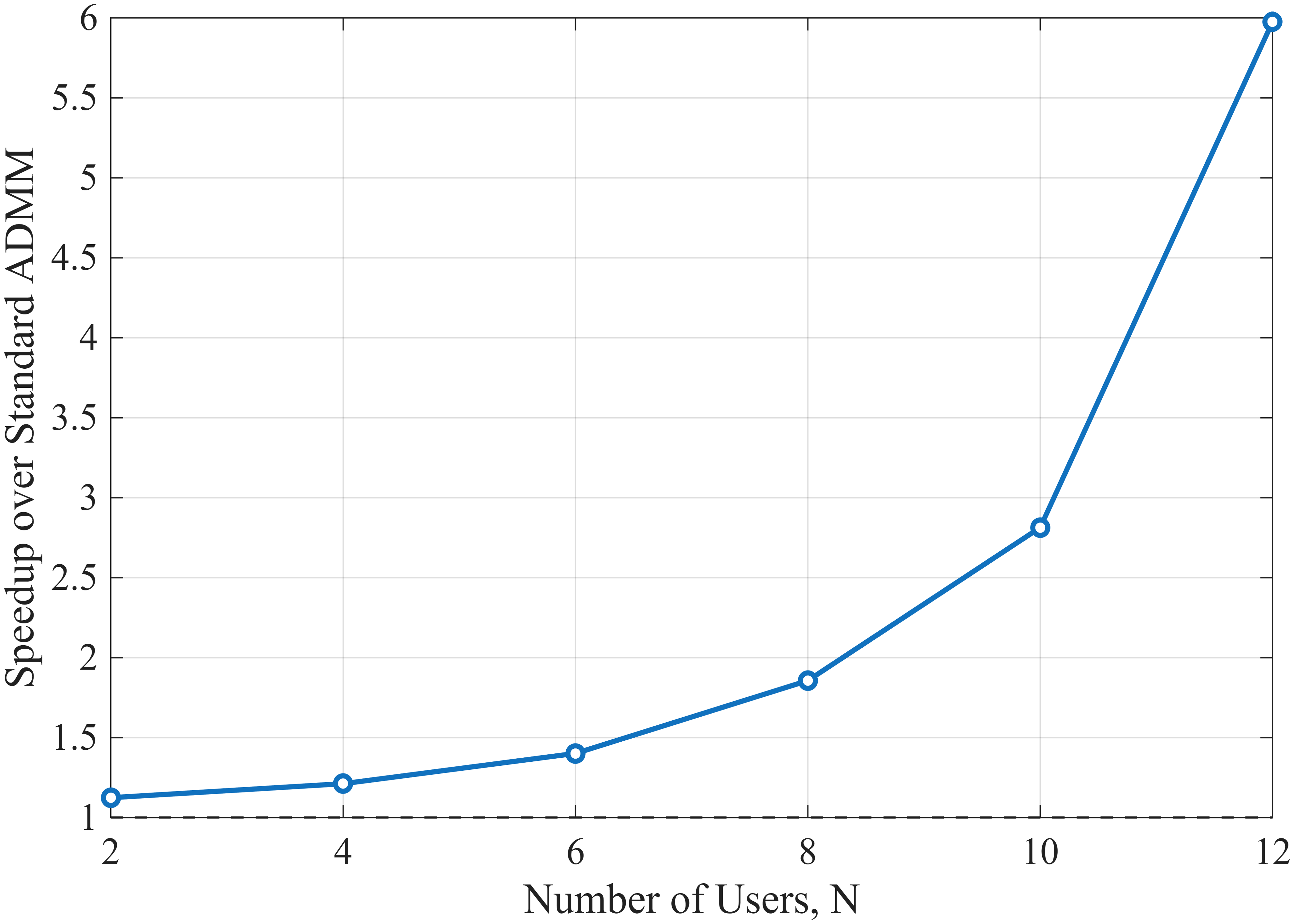}
    \caption{Structural speedup versus $N$ with $M=16$.}
    \label{fig:admm_speedup_N}
\end{subfigure}
\hfill
\begin{subfigure}[t]{0.48\textwidth}
    \centering
    \includegraphics[width=\linewidth]{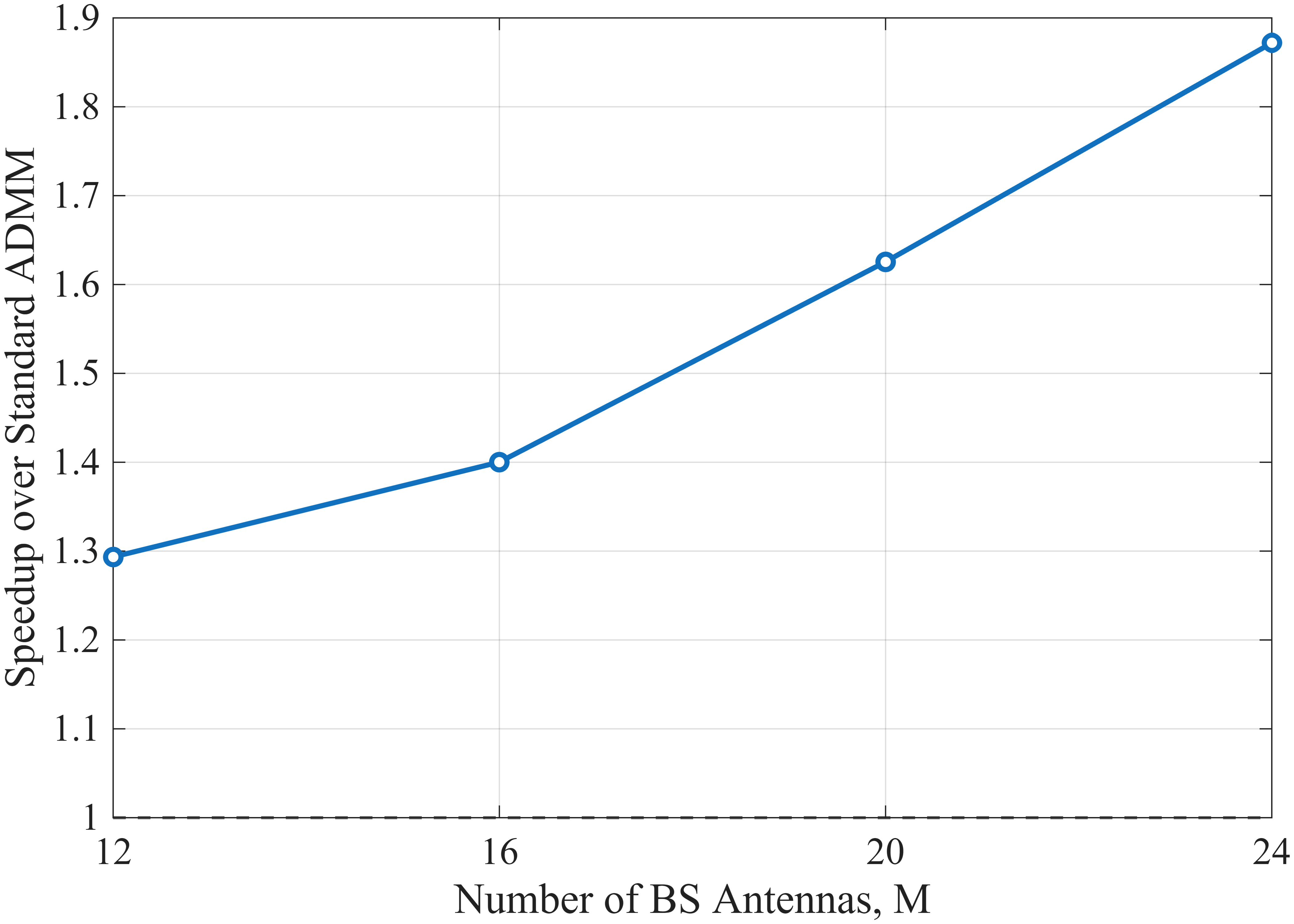}
    \caption{Structural speedup versus $M$ with $N=6$.}
    \label{fig:admm_speedup_M}
\end{subfigure}
\caption{Per-iteration computational speedup of structured ADMM
relative to standard ADMM.}
\label{fig:admm_speedup}
\end{figure*}

To isolate the computational gain of the proposed structural
exploitation from variations in the number of outer SCA iterations,
Fig.~\ref{fig:admm_cost} reports the average computational time per
ADMM iteration. This comparison is particularly relevant since the
standard and structured implementations solve the same consensus
QCQP and differ only in how its algebraic structure is exploited. As shown in Fig.~\ref{fig:admm_cost_N}, the per-iteration cost of
standard ADMM increases rapidly with $N$, whereas the growth of the
structured implementation is considerably slower. This difference
becomes particularly pronounced at large $N$, where the cost of the
standard implementation is dominated by operations in the full
$MN$-dimensional space. In contrast, the proposed implementation
exploits the Kronecker structure of the global quadratic term to
reuse an $M$-dimensional factorization across users and exploits the
rank-one local structure to avoid generic high-dimensional
projections. The antenna-scaling results in Fig.~\ref{fig:admm_cost_M} exhibit
the same trend, although the difference is less pronounced than in
the user-scaling case. With fixed $N$, increasing $M$ enlarges the
beamforming dimension but does not increase the number of
user-specific local updates. By contrast, increasing $N$ enlarges
the stacked dimension and simultaneously introduces additional
user-dependent updates. This explains the substantially stronger
growth of the standard ADMM cost with $N$.

The resulting structural gain is summarized in
Fig.~\ref{fig:admm_speedup}, where the speedup is defined as the
ratio between the average per-iteration times of standard and
structured ADMM. The gain increases with both system dimensions,
but is substantially more pronounced with increasing $N$. For the user scaling in Fig.~\ref{fig:admm_speedup_N}, the speedup
increases from approximately $1.1\times$ at $N=2$ to nearly
$6\times$ at $N=12$. For the antenna scaling in
Fig.~\ref{fig:admm_speedup_M}, the corresponding gain increases
from approximately $1.3\times$ at $M=12$ to $1.9\times$ at
$M=24$. The markedly stronger gain with $N$ is consistent with the
proposed decomposition: increasing the number of users increases
both the dimension of the unstructured consensus problem and the
number of user-dependent updates, whereas the proposed global
factorization remains $M$-dimensional and is reused across users.
These results directly validate the computational scaling predicted
by the complexity analysis and confirm that the end-to-end gains in
Fig.~\ref{fig:cvx_speedup} originate from the proposed structural
exploitation rather than from differences in the outer SCA
convergence.

\begin{figure}[t]
    \centering
    \includegraphics[width=\linewidth]{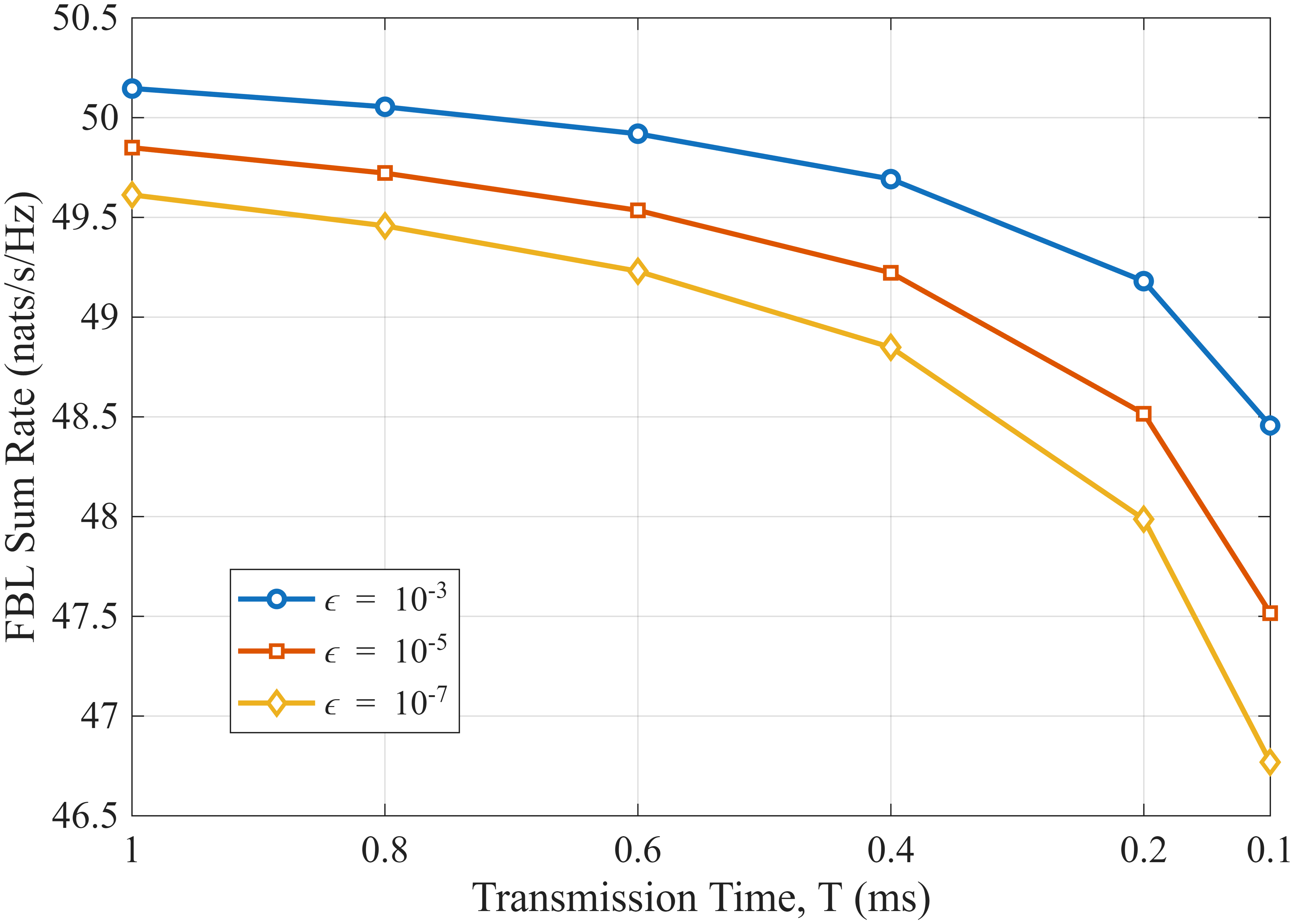}
    \caption{FBL sum rate versus transmission time for different decoding
error probabilities with $M=16$ and $N=8$.}
    \label{fig:fbl_latency}
\end{figure}

To further characterize the URLLC performance,
Fig.~\ref{fig:fbl_latency} illustrates the impact
of transmission time and decoding reliability
on the FBL sum rate, with the beamformers
optimized separately for each considered
configuration using the proposed structured
ADMM-SCA method. The bandwidth is fixed at $1$~MHz, such that reducing
the transmission time from $1$~ms to $0.1$~ms corresponds to decreasing
the blocklength from $1000$ to $100$ channel uses. For all considered
reliability levels, the FBL sum rate decreases as the transmission time
is reduced. This degradation becomes more pronounced in the short-latency
regime due to the increased finite-blocklength penalty associated with
the reduced blocklength. The results also demonstrate the rate--reliability tradeoff. For a
given transmission time, imposing a more stringent decoding error
probability reduces the achievable FBL sum rate. In particular,
$\varepsilon=10^{-7}$ yields a consistently lower sum rate than
$\varepsilon=10^{-5}$ and $\varepsilon=10^{-3}$ over the entire
transmission-time range. This behavior follows from the increase in
$Q^{-1}(\varepsilon)$ as $\varepsilon$ decreases, which increases the
channel-dispersion penalty. Hence, these results illustrate the joint impact of
transmission time and decoding reliability on
the achievable FBL sum rate in latency-critical
cellular IoT networks.

%\FloatBarrier
\section{Conclusion}
\label{sec:conclusion}
This paper developed a structure-exploiting consensus
ADMM-SCA framework for computationally efficient
FBL beamforming in downlink MU-MISO URLLC systems
supporting latency-critical cellular IoT applications. By exploiting the Kronecker structure of the global update and the
rank-one structure of the local QCQP constraints, the proposed method
replaces full $MN$-dimensional operations with a shared
$M\times M$ factorization and rank-one local projection operations. The numerical results showed that the proposed
structured ADMM-SCA closely matches the CVX-SCA benchmark in terms of
the attained FBL sum rate and outer SCA convergence, while requiring
substantially lower computational effort. The computational advantage becomes particularly pronounced as the
number of users increases, since increasing $N$ enlarges both the
beamforming dimension and the number of user-dependent constraints.
The comparison with standard ADMM-SCA further shows that this gain
cannot be attributed to the use of ADMM alone, but results from
exploiting the specific algebraic structure of the FBL beamforming
problem. For increasing $M$ with a fixed number of users, the
computational growth is more moderate for both ADMM implementations,
while the proposed structured solver consistently retains a lower
per-iteration cost. These observations identify user scaling as the
more demanding computational regime and demonstrate
the benefit of structure-aware optimization for
serving an increasing number of latency-critical
cellular IoT nodes. Together with the observed impact of transmission
time and decoding reliability on the achievable
FBL sum rate, these results support computationally
efficient FBL-aware radio-resource management under
stringent URLLC service requirements.

\bibliographystyle{IEEEtran}
%\bibliography{references}
\bibliography{references_abbreviated}

\end{document}